\documentclass[twocolumn, showkeys, superscriptaddress, amsmath, amssymb,
 aps, pra, floatfix, nofootinbib]{revtex4-1}

\usepackage{graphicx}
\usepackage{dcolumn}
\usepackage{bm}
\usepackage{hyperref}
\usepackage{xcolor,tikz, circuitikz}
\usepackage{easyReview}
\usepackage{float}
\usepackage{multirow}

\hypersetup{
colorlinks,
linkcolor={blue},
citecolor={blue},
urlcolor={blue!80!black}
}

\usepackage{amsthm}

\usetikzlibrary{positioning, shapes.geometric}

\newtheorem{definition}{Definition}

\newcommand{\mean}[1]{\langle#1\rangle}
\newcommand{\Log}[1]{\log\left(#1\right)}

\begin{document}

\title{Evaluating Quantumness, Efficiency and Cost of\\ Quantum Random Number Generators via Photon Statistics}

\homepage[This is a substantially updated and extended version of the paper entitled ``The Two Sides of Photon Arrival-time-based QRNGs'' presented in 1$^{st}$ International Conference on Emerging Trends in Optical Technologies (ETOT-I) held in SRM University, Andhra Pradesh, India on January 2-4, 2025.  In this extended journal version, Sections~1, 2, 3, 4, 5 have been thoroughly rewritten with added explanations, new equations and results. Sections~1.1, 3.4, 6, and Appendices~A, B, C, D, E are completely new additions. Further, Figures~1-3, 5-7, 9-14 have been newly added with relevant discussions.]{}

\author{Goutam Paul}
\email{goutam.paul@isical.ac.in}
\affiliation{Cryptology and Security Research Unit, 
Indian Statistical Institute,
Kolkata 700108, India}

\author{Nirupam Basak}
\email{nirupambasak2020@iitkalumni.org}
\affiliation{Cryptology and Security Research Unit,
Indian Statistical Institute,
Kolkata 700108, India} 

\author{Soumya Das}
\email{s.das2@tue.nl}
\affiliation{Dept. of Math. and Comp. Sci., Eindhoven University of Technology, Eindhoven, Netherlands}

\begin{abstract}
This work presents two significant contributions from the perspectives of quantum random number generator (QRNG) manufacturers and users. For manufacturers, the conventional method of assessing the quantumness of single-photon-based QRNGs through mean and variance comparisons of photon counts is statistically unreliable due to finite sample sizes. Given the sub-Poissonian statistics of single photons, confirming the underlying distribution is crucial for validating a QRNG's quantumness. We propose a more efficient two-fold statistical approach to ensure the quantumness of optical sources with the desired confidence level. Additionally, we demonstrate that the output of QRNGs from exponential and uniform distributions exhibit similarity under device noise, deriving corresponding photon statistics and conditions for $\epsilon$-randomness.

From the user's perspective, the fundamental parameters of a QRNG are quantumness, efficiency (entropy and random number generation rate), and cost. Our analysis reveals that these parameters depend on three factors, namely, expected photon count per unit time, external reference cycle duration, and detection efficiency. A lower expected photon count enhances entropy but increases cost and decreases the generation rate. A shorter external reference cycle boosts entropy but must exceed a minimum threshold to minimize timing errors, with minor impacts on cost and rate. Lower detection efficiency enhances entropy and lowers cost but reduces the generation rate. Finally,  to validate our results, we perform statistical tests like NIST, Dieharder, AIS-31, ENT etc. over the data simulated with different values of the above parameters. Our findings can empower manufacturers to customize QRNGs to meet user needs effectively.
\end{abstract}

\maketitle

\vspace{2pc}

\section{Introduction}

Random numbers play a vital role in various applications, including Monte Carlo simulations, software development, physics and chemistry experiments, cryptography, and commercial uses like slot machines and lotteries~\cite{RevModPhys.87.1067,MonteCarlo,PhysicsPhysiqueFizika.1.195,Bennett,Crypto,konstantinou2004electronic,yamanaka2017enumeration}. In cryptography, secure communication relies on high-quality randomness. Classical pseudo-random number generators (PRNGs), being deterministic, lack true unpredictability. In contrast, true random number generators (TRNGs) derive randomness from physical entropy sources like quantum effect, noise, free oscillations, chaos, etc.~\cite{Schindler2009,Stipčević2014, MA2024115310, ZHAO2023113407, Ma_2025}.

Classical TRNGs rely on the unpredictability of physical processes due to incomplete knowledge~\cite{Stipčević2014,mannalatha2023comprehensive}, but future research may render these processes predictable. Quantum TRNGs (QRNGs), however, leverage quantum mechanics to generate fundamentally unpredictable numbers~\cite{ma2016quantum, Zhang2023}, ensuring lasting randomness.

Different technologies, including photon arrival time~\cite{10.1063/1.2720728,10.1063/1.2961000,WayneRNG,WayneQRNG,10.1063/1.3578456,10.1063/1.4863224,7270265,zhang2021simple, solymos2024correlation}, tunneling~\cite{bernardo2017extracting,10.1063/5.0055955}, photon detection~\cite{acsomega,wang2025efficient}, fluctuation~\cite{gabriel2010generator,PhysRevApplied.3.054004,Guo:19,ZQVR,10.1063/1.5078547,gehring2021homodyne}, phase noise~\cite{Qi:10,PhysRevE.81.051137,PhysRevA.81.063814,Xu:12,10.1063/1.4922417,Yang:16,app10072431}, etc. have been proposed to produce QRNGs~\cite{RevModPhys.89.015004,mannalatha2023comprehensive}. Several manufacturers such as ID Quantique~\cite{quantis}, Qutools~\cite{quRNG}, QNuLabs~\cite{TROPOS}, QuintessenceLabs~\cite{qStream}, Mars Innovation~\cite{quantoss}, Quantum Computing Inc.~\cite{uQRNG}, Quside~\cite{quside}, QuantumCTek~\cite{QCTek} are producing QRNGs following these technologies. These QRNGs are commercially available as off-the-shelf black-box devices. As the photon arrival-time-based QRNGs are the most general and most of companies follow this approach, in this work we focus on the analysis of this type of QRNGs only.

Since no black-box randomness-testing algorithm can distinguish between a pseudorandom number generator (PRNG) and a quantum random number generator (QRNG)~\cite{PhysRevA.109.022243}, the first step in validating the \emph{quantumness} of a random number generator is to gather detailed information about the intermediate components, such as attenuator and detector parameters, and photon statistics. This data may then enable a strategy to determine whether the device operates in the classical or quantum regime. Here, by quantumness we denote whether the underlying photon-counting-statistics follows sub-Poissonian distribution ($\text{mean} > \text{variance}$). For example, for a probability distribution given by $P(n)=\frac{e^{-\alpha\mu}(\alpha\mu)^{\alpha n}}{(\alpha n)!},\;\text{if }\alpha n\text{ is integer, and }0$ otherwise, for $n=0,1,2,\dots$, where $\mu,\alpha$ are some positive real numbers, the mean would be $\mu$ and the variance would be $\frac{\mu}{\alpha}$. Thus it denotes a sub-Poissonian, Poissonian, or super-Poissonian distribution for $\alpha>1,\alpha=1,\alpha<1$, respectively. Therefore, if $\alpha>1$, we can surely say that this falls under the quantum regime. We provide real-world examples in~\ref{sec:g2t}.

Numerous photon-number-resolving detectors have been proposed~\cite{PhysRevA.71.033819,Divochiy2008,Kardyna2008,PhysRevA.99.043822,Natarajan2012,Schuck2013,Hadfield2009,9844820,Provaznik:20,Marsili2009} and experimentally demonstrated~\cite{1263782,PhysRevA.67.061801,10.1063/1.1596723,PhysRevA.71.061803,PhysRevA.75.062325,Cheng2023,Lita:08,Calkins:13,Reddy:20,Marsili2013,10.1063/1.5000001,Zhang2017,Pernice2012,doi:10.1021/acsphotonics.4c01680,10.1063/1.1388868} in recent years, providing photon-counting statistics for analysis of the underlying distribution. In most analyses of photon arrival-time-based QRNGs~\cite{Jones:23,Perina:24,Perina24,PhysRevA.109.043505}, photon statistics are examined by comparing the sample mean and variance or using the Mandel parameter~\cite{Mandel:79}. For instance, All\'{e}aume \emph{et al.}~\cite{Alléaume_2004} analyzed photon count data using the Mandel parameter~\cite[Eq.~(11a)]{Mandel:79}, concluding that its negative value, which indicates sub-Poissonian statistics, signifies non-classical photon sources. While analytically valid, this approach may be misleading for finite samples. A negative Mandel parameter can appear even for Poissonian sources like coherent light due to finite-size sampling error, where the sample mean and variance deviate from those of the true distribution~\cite{BERG2005865}. Thus, a single parameter value from a finite dataset is insufficient to infer the underlying photon statistics.

Another method to assess the quantumness of light is through photon antibunching~\cite{RevModPhys.54.1061}. Although antibunched photons indicate non-classicality, the absence of antibunching does not imply classical behavior; non-classical light may exhibit photon bunching~\cite{PhysRevA.41.475,fox2006quantum}. This criterion also relies on statistical equality, making it similarly inconclusive in certain cases. A detailed discussion appears in Section~\ref{sec:g2t}.

On the other hand, a single-photon source ensures the quantumness of a QRNG~\cite{RevModPhys.54.1061,fox2006quantum}, though such sources remain experimentally challenging~\cite{PhysRevLett.72.210,kim1999single}. Various approaches have been proposed to generate single photons~\cite{PhysRevLett.83.2722,lounis2000single,PhysRevLett.86.1502,santori2002indistinguishable,10.1063/1.1577828,PhysRevLett.89.067901,adam2024single,wang2024multimode}. 

Time-bin-based QRNGs typically use attenuated coherent sources to approximate single-photon emission and generate random numbers with exponential~\cite{WayneQRNG,WayneRNG,10.1063/1.3578456} or uniform~\cite{10.1063/1.4863224} distributions under ideal conditions. However, in practice, deviations occur. For example, Nie \emph{et al.}~\cite{10.1063/1.4863224} demonstrated that uniformity, and thus min-entropy, depends on parameters like photon numbers and detection efficiency. This highlights the importance of analyzing photon statistics under realistic conditions.

Although QRNGs aim to produce uniformly distributed outputs, imperfections in hardware introduce deviations. As a result, classical post-processing is required to extract uniform randomness. It is important to note that such processing does not increase entropy, but compresses the raw output to extract existing min-entropy, often at the cost of reduced generation rates.

\subsection{\label{sec:motiv}Motivation}

Given that both PRNGs and QRNGs produce sequences of 0's and 1's, merely analyzing these sequences is not sufficient to definitively differentiate between the two~\cite{PhysRevA.109.022243}. Statistical tests like NIST~\cite{10.5555/2206233}, Dieharder~\cite{Die}, AIS-$31$~\cite{killmann2011proposal}, ENT~\cite{ENT} etc. are available to test the randomness of an RNG. However, from the results of these tests, it is not possible to say whether an RNG is quantum or not. So a pertinent question arises: 

\emph{\textbf{Q1: } How do we test whether an alleged QRNG is truly quantum or not?}

Several device-independent QRNG (DI-QRNG) schemes~\cite{pironio2010random,10.1145/2885493,acin2016certified,bierhorst2018experimentally,liu2018device,PhysRevLett.120.010503,9153851} and certified randomness protocols~\cite{pironio2010random,6326426,Passaro_2015,acin2016certified,Nguyen:18,10.1093/oso/9780198788416.003.0008,PhysRevX.10.041048,zhang2021simple,Leone:23,PhysRevLett.133.020802,PRXQuantum.5.020348} based on Bell violation and non-separability~\cite{acin2016certified,Alicki_2008,Alicki_2008a,Martin_2023} have been proposed. These approaches do not rely on trusted devices or require post-processing, but they demand costly quantum resources such as entangled qubits. Recent efforts~\cite{Mironowicz_2016,10.1145/3564246.3585145} explore locally certified randomness without entanglement, though they still require random seeds and bootstrapping steps, functionally similar to classical post-processing.

In contrast, optical trusted-device QRNGs~\cite{mannalatha2023comprehensive} assume that both the source and detection setup are well-characterized and isolated from external influence. While less robust than DI-QRNGs, they are more practical and can be verified using photon-count statistics~\cite{PhysRevA.41.475,fox2006quantum,RevModPhys.68.127,qute.202100062}, offering enhanced security under trusted assumptions. A detailed comparison is provided in Section~\ref{sec:DIQRNG}. This work focuses on optical trusted-device QRNGs, and introduces a two-fold statistical approach to verify their quantumness.

Side-channel attacks~\cite{8924447,8942152,chowdhury2021physical} may target the post-processing stage to extract information from the output. Hence, minimizing post-processing is essential to reduce compression overhead, improve throughput, and mitigate attack surfaces. For this, the raw output should be as close to uniform as possible. Nie \emph{et al.}~\cite{10.1063/1.4863224} derived a lower bound on the min-entropy of raw output under device imperfections. Based on this min-entropy, suitable post-processing can be applied to retain entropy from the quantum source. As all practical systems involve errors, this raises a critical question:

\emph{\textbf{Q2: }How to increase the min-entropy of the raw data in presence of the device errors?}

Q1 and Q2 regarding quantumness and randomness are crucial from the manufacturer's perspective. Besides these, the random number generation rate and the cost of the QRNG are also important parameters from the user's perspective. To achieve a comprehensive treatise, we analyze QRNG from both perspectives. Depending on the users' requirements, the trade-off between the randomness of the quantum source, the random number generation rate, and the cost of the QRNG can vary. For example, in the case of secure communication, the randomness from the quantum source should have the highest priority, and also the generation rate should be high. On the other hand, in token generation or lottery games, a low rate of random number generation would be sufficient. Therefore, to fulfill the user's requirement, the answers to the following questions are required.

\emph{\textbf{Q3: }How to choose the optical parameters (a) the photon numbers in unit time, (b) the time cycle and (c) the detection efficiency of a QRNG depending on the user's requirements?}

\subsection{Our Contributions}

In this paper, we aim to evaluate the quantumness of photon arrival-time-based QRNG from the manufacturer's perspective as answers to Q1 and Q2 and randomness, random number generation rate, and cost from the user's perspective as the answer to Q3. A similar analysis can also be performed for other QRNG models. For the manufacturer, we first address the validation of quantum sources using photon counting statistics. Considering that the sample data for the number of photons within a fixed time is available, we suggest a two-fold statistical method to verify whether the source of the sample is quantum or not, overcoming the limitations of the existing approaches discussed above. The method consists of interval estimation and hypothesis testing to ensure sub- or super-Poissonian distribution up to a desired confidence level. Several photon-number-resolving detectors have been proposed~\cite{PhysRevA.71.033819,Divochiy2008,Kardyna2008,PhysRevA.99.043822,Natarajan2012,Schuck2013,Hadfield2009,9844820,Provaznik:20,Marsili2009} and demonstrated~\cite{1263782,PhysRevA.67.061801,10.1063/1.1596723,PhysRevA.71.061803,PhysRevA.75.062325,Cheng2023,Lita:08,Calkins:13,Reddy:20,Marsili2013,10.1063/1.5000001,Zhang2017,Pernice2012,doi:10.1021/acsphotonics.4c01680,10.1063/1.1388868} in recent years that can be used to perform these tests.

As another contribution, also from the manufacturer's perspective, we derive the photon statistics for two photon arrival-time-based QRNGs in practical scenarios. We consider power instability as the error of the source and the attenuators and detection inefficiency as the error of a single photon detector (SPD).

From the user's perspective, we see that the maximum entropy from the quantum source may not be achieved due to some practical constraints related to the random number generation rate, the timing error in registering the photon detection time and the cost of the QRNG. In particular, we show the following.
\begin{itemize}
\item If the expected number of photons within a unit time is small, the min-entropy from the quantum source would be high. However, it would increase the cost of the QRNG, and it would also reduce the random number generation rate.
\item Although a small external reference cycle of the QRNG increases the min-entropy from the quantum source, to minimize the timing error, it should be bounded by a specific value. This reference cycle does not have much effect on the cost or random number generation rate of the QRNG.
\item A detector with less efficiency increases the min-entropy from the quantum source and also reduces the cost of the QRNG. However, it also reduces the random number generation rate.
\end{itemize}

Finally, we simulated raw data depending on the photon statistics we have derived and performed statistical tests like NIST~\cite{10.5555/2206233}, Dieharder~\cite{Die}, AIS-$31$~\cite{killmann2011proposal} and ENT~\cite{ENT} on those raw data. We show that these statistical tests support our theoretical deduction regarding randomness from quantum sources. Nie \emph{et al.}~\cite{10.1063/1.4863224} performed NIST tests after post-processing and showed that the post-processed data passed all of the 15 tests from the NIST suite.

This paper is structured as follows. In Section~\ref{sec:coh_pulse} we explore coherent pulse-based QRNGs and mention their weakness. Section~\ref{sec:2fold} reveals our two-fold proposal to check the quantumness of a photon-based quantum device as an answer to Q1. Detailed discussion on coherent pulse-based QRNGs has been performed in Section~\ref{sec:muTd} to answer Q2. In this section, we have also mentioned how our findings are different from the work of Nie \emph{et al.}~\cite{10.1063/1.4863224}. Section~\ref{sec:constraint} answers Q3 by separately analyzing the effects of the parameters mentioned in that question. The answer to Q3 by combining the randomness with random number generation rate and entropy for maximum rate is presented in Section~\ref{sec:eval_simult}. Results from different randomness testing suites for our simulated random numbers based on the analysis of Section~\ref{sec:muTd} are given in Section~\ref{sec:NIST}. Finally, in Section~\ref{sec:conc} we conclude our work.

\section{\label{sec:coh_pulse}Existing QRNG Models based on Coherent Pulse}

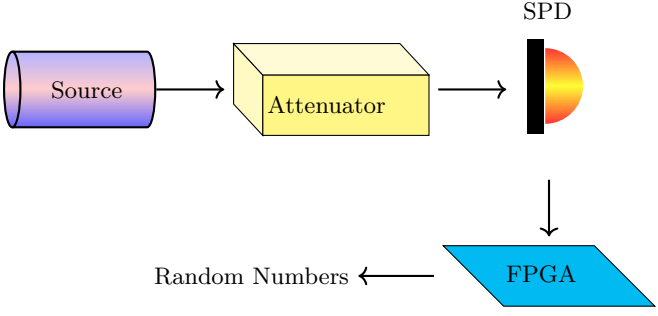
\begin{figure}[htpb]
\centering
\begin{tikzpicture}
\node[cylinder, minimum height = 2cm, minimum width = 1cm, shape aspect=.5, top color=blue!30, bottom color=blue!60, middle color=red!20, shape border rotate=180, draw, thick](source) {Source};
\draw[thick, ->] (source.east) --+(0.9, 0) node[anchor = west] (Att) {
\begin{tikzpicture}
\pgfmathsetmacro{\cubex}{2.2}
\pgfmathsetmacro{\cubey}{0.8}
\pgfmathsetmacro{\cubez}{1}
\draw[fill=yellow!60] (0,0,0) -- ++(\cubex,0,0) -- ++(0,-\cubey,0) -- ++(-\cubex,0,0) -- cycle ;
\draw[fill=yellow!30] (0,0,0) -- ++(0,\cubey,\cubez) -- ++(0,-\cubey,0) -- ++(0,-\cubey,-\cubez) -- cycle;
\draw[fill=yellow!20] (0,0,0) -- ++(\cubex,0,0) -- ++(0,\cubey,\cubez) -- ++(-\cubex,0,0) -- cycle;
\node at (-0.05,-0.4,0) {Attenuator};
\end{tikzpicture}};
\draw[thick, ->] (Att.east) --+(0.9, 0) node[anchor = west] (SPD) {
\begin{tikzpicture}
\node[semicircle, shape border rotate = 270, top color = red!80, bottom color = red!80, middle color = yellow!80, minimum height = .5cm] (SPD) {};
\node[rectangle, left = 0cm of SPD, minimum height = 1.25cm, fill = black, aspect = 0.5]{};
\node at (-0.4,1) {SPD};
\node at (0,-1.2) {};
\end{tikzpicture}
};
\draw[thick, ->] (SPD.mid) --+(0, -0.75) node[anchor = north] (FPGA) {
\begin{tikzpicture}
\draw[fill = cyan!80] (0, 0) -- ++(2, 0) -- ++(-0.8, 0.8) -- ++(-2, 0) -- cycle;
\tikzset{trapezium stretches=true}
\node at (0.5, 0.65) {FPGA};
\end{tikzpicture}
};
\draw[thick, ->] (FPGA.west) --+(-1, 0) node[anchor = east] {Random Numbers};
\end{tikzpicture}
\caption{A basic block diagram of the photon arrival time-based QRNG using weak coherent pulses. Coherent pulses, coming from the source, are sent through attenuators to reduce their power so that the probability of having multiple photons within a predefined time interval becomes negligible~\cite{argillander2023quantum}. A SPD detects these photons and the detection time (photon arrival time) has been registered. The Field Programmable Gate Array (FPGA) provides random numbers based on the registered time stamps.}
\label{fig:block-diagram}
\end{figure}

Although a perfect single photon source ensures the quantum behaviour of light, it is well known that producing single photon states is experimentally challenging~\cite{PhysRevLett.72.210,kim1999single}. However, different experimental approaches~\cite{PhysRevLett.83.2722,lounis2000single,PhysRevLett.86.1502,santori2002indistinguishable,10.1063/1.1577828,PhysRevLett.89.067901,adam2024single,wang2024multimode} have been performed to produce single photons. One of the approaches is using a coherent photon source followed by attenuators. This approach produces photon pulses with negligible probability of having multiple photons within a time interval~\cite{argillander2023quantum}. In this process, the photons follow Poissonian statistics. A generic block diagram of the photon arrival time-based QRNG using photons from coherent sources is given in Fig.~\ref{fig:block-diagram}.

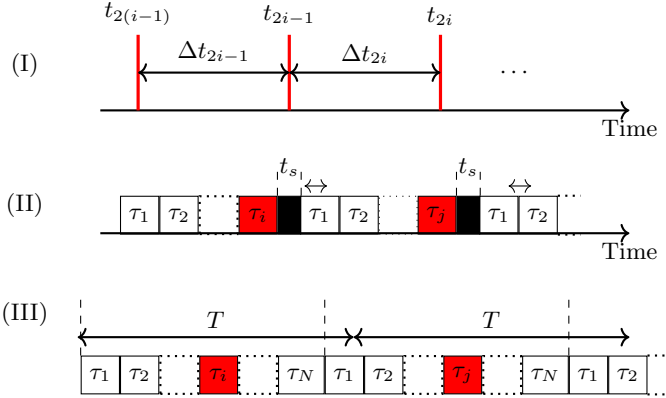
\begin{figure}[htpb]
\centering
\begin{tikzpicture}
\node[rectangle, draw, minimum size = .5cm] (1) {$\tau_1$};
\node[rectangle, draw, right = 0cm of 1, minimum size = .5cm] (2) {$\tau_2$};
\node[rectangle, thick, draw, dotted, right = 0cm of 2, minimum size = .5cm] (3) {};
\node[rectangle, fill = red, draw, right = 0cm of 3, minimum size = .5cm] (4) {$\tau_i$};
\node[rectangle, thick, draw, dotted, right = 0cm of 4, minimum size = .5cm] (5) {};
\node[rectangle, draw, right = 0cm of 5, minimum size = .5cm] (6) {$\tau_N$};
\node[rectangle, draw, right = 0cm of 6, minimum size = .5cm] (7) {$\tau_1$};
\node[rectangle, draw, right = 0cm of 7, minimum size = .5cm] (8) {$\tau_2$};
\node[rectangle, thick, draw, dotted, right = 0cm of 8, minimum size = .5cm] (9) {};
\node[rectangle, fill = red, draw, right = 0cm of 9, minimum size = .5cm] (10) {$\tau_j$};
\node[rectangle, thick, draw, dotted, right = 0cm of 10, minimum size = .5cm] (11) {};
\node[rectangle, draw, right = 0cm of 11, minimum size = .5cm] (12) {$\tau_N$};
\node[rectangle, draw, right = 0cm of 12, minimum size = .5cm] (13) {$\tau_1$};
\node[rectangle, draw, right = 0cm of 13, minimum size = .5cm] (14) {$\tau_2$};
\draw[dashed] (1.north west) -- ++(0, 0.75) (6.north east) -- ++(0, 0.75) (12.north east) -- ++(0, 0.75);
\draw[<->, thick] (-0.3, 0.5) -- ++(3.65,0) node[midway, above] {$T$};
\draw[<->, thick] (3.35, 0.5) -- ++(3.65, 0) node[midway, above] {$T$};
\draw[dotted, thick] (14.north east) -- ++(0.3,0);
\draw[dotted, thick] (14.south east) -- ++(0.3,0);
\node at (-1.0, .8) (III) {(III)};

\node[rectangle, draw, minimum size = .5cm, above = 1.6cm of 2] (21) {$\tau_1$};
\node[rectangle, draw, right = 0cm of 21, minimum size = .5cm] (22) {$\tau_2$};
\node[rectangle, thick, draw, dotted, right = 0cm of 22, minimum size = .5cm] (23) {};
\node[rectangle, fill = red, draw, right = 0cm of 23, minimum size = .5cm] (24) {$\tau_i$};
\node[rectangle, draw, right = 0.3cm of 24, minimum size = .5cm] (25) {$\tau_1$};
\node[rectangle, draw, right = 0cm of 24, minimum height = .5cm, minimum width = 0.3cm, fill] (0) {};
\node[above = 0.15cm of 0] {$t_s$};
\node[rectangle, draw, right = 0cm of 25, minimum size = .5cm] (26) {$\tau_2$};
\node[rectangle, draw, dotted, right = 0cm of 26, minimum size = .5cm] (27) {};
\node[rectangle, fill = red, draw, right = 0cm of 27, minimum size = .5cm] (28) {$\tau_j$};
\node[rectangle, draw, right = 0.3cm of 28, minimum size = .5cm] (29) {$\tau_1$};
\node[rectangle, draw, right = 0cm of 28, minimum height = .5cm, minimum width = 0.3cm, fill] (00) {};
\node[above = 0.15cm of 00] {$t_s$};
\node[rectangle, draw, right = 0cm of 29, minimum size = .5cm] (30) {$\tau_2$};
\draw[dotted, thick] (30.north east) -- ++(0.3,0);
\draw[dotted, thick] (30.south east) -- ++(0.3,0);
\draw[thick, ->] (0, 1.875) -- ++(7, 0) node[below] {Time};
\draw[<->] (2.7, 2.5) -- ++(0.3, 0);
\draw[<->] (5.4, 2.5) -- ++(0.3, 0);
\draw[dashed] (0.north west) -- ++(0, .4);
\draw[dashed] (0.north east) -- ++(0, .4);
\draw[dashed] (00.north west) -- ++(0, .4);
\draw[dashed] (00.north east) -- ++(0, .4);
\node[above = 0.9cm of III] {(II)};

\draw[thick, ->] (0, 3.5) -- ++(7, 0) node[below] {Time};
\draw[very thick, red] (0.5, 3.5) -- ++(0, 1) node[above, black] {$t_{2(i-1)}$};
\draw[very thick, red] (2.5, 3.5) -- ++(0, 1) node[above, black] {$t_{2i-1}$};
\draw[very thick, red] (4.5, 3.5) -- ++(0, 1) node[above, black] {$t_{2i}$};
\node at (5.5, 4) {$\cdots$};
\draw[thick, <->] (0.5, 4) -- (2.5, 4) node[midway, above] {$\Delta t_{2i-1}$};
\draw[thick, <->] (2.5, 4) -- (4.5, 4) node[midway, above] {$\Delta t_{2i}$};
\node[above = 2.75cm of III] {(I)};
\end{tikzpicture}
\caption{Time diagram for the arrivals of the photons. There are three different methods to generate random numbers from photon arrival times. (I) Photons are detected in a continuous time reference at time $\{t_0,t_1,\dots t_{2(i-1)},t_{2i-1},t_{2i},\dots\}$. The $i$-th bit of the random number will be $0$ if $\Delta t_{2i-1}<\Delta t_{2i}$ and $1$ if $\Delta t_{2i-1}>\Delta t_{2i}$. (II) Time is divided into bins of equal length. If a photon has been detected in the $i$-th bin then $i-1$ will be the corresponding random number. A new cycle of bins starts after a fixed sleep time ($t_s$) of the SPD. (III) An external time reference of cycle $T$ has been used. Each cycle is divided into $N$ equal time-bins $\tau_1,\tau_2,\dots,\tau_N$. For each cycle, if the first photon has been detected in $i$-th bin then $i-1$ will be the corresponding random number. SPD will be in sleep mode for the remaining time of the cycle.}
\label{fig:time_diag}
\end{figure}

Several methods exist for generating random numbers from registered photon arrival times, illustrated in Fig.~\ref{fig:time_diag}. In Fig.~\ref{fig:time_diag}(I), given arrival times $\{t_0, t_1, \dots, t_{2i}\}$, two consecutive non-overlapping time intervals $\Delta t_{2i-1} = t_{2i-1} - t_{2i-2}$ and $\Delta t_{2i} = t_{2i} - t_{2i-1}$ are used to generate the $i$-th bit as:
\begin{equation}
\begin{cases}
0, & \text{if } \Delta t_{2i-1} < \Delta t_{2i}, \\
1, & \text{if } \Delta t_{2i-1} > \Delta t_{2i}.
\end{cases}
\end{equation}
This method~\cite{10.1063/1.2961000, Schranz2024} has low efficiency, producing one bit per two photon detections.

Faster methods have been proposed~\cite{10.1063/1.2720728,WayneQRNG,WayneRNG,10.1063/1.3578456,10.1063/1.4863224, solymos2024correlation}. In Fig.~\ref{fig:time_diag}(II), time is divided into equal-length bins. If a photon is detected in the $i$-th bin, the value $i-1$ is taken as the random number. A new cycle begins after a fixed detector sleep time. Since photon detection follows a Poisson process, the probability that the waiting time $t_w$ between two photons exceeds $t$ is $P(t_w > t)=P(\text{no photon in }(0,t])=P(0,t) = e^{-\mu t}$, indicating an exponential distribution of outcomes~\cite{WayneQRNG,WayneRNG,10.1063/1.3578456,10.1063/1.4863224}.

A third approach~\cite{10.1063/1.2720728}, shown in Fig.~\ref{fig:time_diag}(III), uses an external clock divided into cycles of length $T$, smaller than the detector's dead time. Each cycle is further divided into $N$ equal bins $\tau_1, \tau_2, \dots, \tau_N$. If the first detected photon in a cycle arrives in bin $\tau_i$, the output is $i-1$. The detector remains inactive for the rest of the cycle.

Let $n$ photons arrive within time $T$ after attenuation. Since the photon count follows Poisson distribution~\eqref{E:Psn}, the probability that all $n$ photons fall in bin $\tau_i$ is:
\begin{align}
P(i;\text{all}|n) &= \frac{P(n, |\tau_i|)\, P(0, \frac{i-1}{N}T)\, P(0, T - \frac{i}{N}T)}{P(n, T)} \notag \\
&= \frac{e^{-\mu |\tau_i|} (\mu |\tau_i|)^n / n! \cdot e^{-\mu \frac{i-1}{N}T} \cdot e^{-\mu(1 - \frac{i}{N})T}}{e^{-\mu T} (\mu T)^n / n!} = \frac{1}{N^n}, \label{E:ideal}
\end{align}
where $|\tau_i| = \frac{T}{N}$. Thus, if exactly one photon is present within $T$ and detection is ideal, it falls in any bin with equal probability $1/N$, producing uniformly distributed outcomes.

However, this ideal case assumes perfect single-photon input and detection, conditions rarely met in practice. In the following, we analyze photon statistics under more realistic experimental scenarios.

\section{\label{sec:2fold}Our Proposal of Two-fold Statistical Method for Testing Quantumness}

For the Poissonian distribution, the mean photon count is the same as its variance. However, it is interesting to note that \emph{mean $=$ variance} does not always imply that the distribution is Poisson. For example, consider the probability distribution
\begin{equation}
\label{E:geo}
P(k)=(1-p)^{k-1}p,k=1,2,3,\dots,
\end{equation}
for $0\leq p\leq1$. If $p=\frac{1}{2}$, this distribution also has mean = variance $=2$. But it is not Poisson.

Also, in experimental methods, the means and variances are calculated from finite data samples. The fineness of the sample cannot perfectly reflect the mean and variance of the underlying distribution. In such a scenario, direct comparison between the mean and the variance may not always work. Suppose we are given a device to test whether the photons emitted from it are in the quantum regime. There are two possibilities -- either the device is a single photon source with sub-Poissonian distribution or it is a source of coherent pulses with Poissonian distribution\footnote{very rarely it will be super-Poissonian}. In the former case, the old method of point estimation will successfully predict the correct nature of the device. However, in the latter case, it will fail most of the time. This may lead to a wrong conclusion about the underlying distribution. For example, in the case of a device based on coherent pulse, which follows a Poisson distribution, a small error in the sample mean or in the sample variance may lead to a conclusion that the underlying distribution is sub-Poissonian, indicating a proper quantum source of light.

This motivates us to propose a \emph{two-fold method} to test the quantumness of the photon source, as an answer to question Q1, whether the source is a purely single-photon source or a source of coherent pulses. First, a statistical test is performed to verify whether the mean is equal to or less than or greater than the variance. Note that any statistical validation of equality is not exact, but rather approximate in practice. Thus, if the above test validates the mean and the variance to be equal, we propose to perform a second test to validate that the distribution is indeed Poisson. The statistical tools used in this section is discussed in~\ref{sec:stat_tools}.

\subsection{First Phase: Interval Estimation}

Since Poissonian and super-Poissonian statistics are explainable using classical theory, only sub-Poissonian distributions can guarantee the quantumness of a quantum device.  We suggest a procedure to decide whether the underlying distribution is sub-Poissonian or not by estimating the mean and the variance of the underlying distribution. However, estimation of the values, known as \emph{point estimation}, is not always exact~\cite{ross2009introduction, mood1974}. Thus, estimating two parameters, the mean and the variance, would increase the error. Therefore, we suggest estimation of an interval for one parameter with some confidence, called \emph{significance level}. Then we can compare the other parameter with this interval.

First, we compare the variance with the interval of the mean. Here, we find an interval where the distribution mean would lie up to some given confidence and a value estimating the variance of the distribution. Then, we compare the estimated variance with the interval to decide the nature of the distribution. Suppose the number of photon counts within a fixed time is given as a sample of size $n$. Then by the central limit theorem, $\frac{\mean X-m}{\sigma/\sqrt{n}}$ is approximately standard normal for large $n$~\cite{hogg2019introduction}, where $\mean X$ is the random variable corresponding to the sample mean $\mean{x}$, and $m,\sigma^2$ are mean and variance of the actual photon-count distribution. Let the sample variance be given by $S^2$. Then
\begin{equation}
\label{E:est_var}
s^2=\frac{n}{n-1}S^2
\end{equation}
is an unbiased and consistent estimate of $\sigma^2$~\cite{hogg2019introduction}. Therefore, by replacing $\sigma$ by $s$, we can say that $Y=\frac{\mean{X}-m}{s/\sqrt{n}}$ is approximately standard normal for large $n$.

Let $1-\epsilon$ be the confidence level. Suppose $y_\epsilon$ is such that 
\begin{equation}
\label{E:est}
P(-y_\epsilon<Y<y_\epsilon)\approx\frac{1}{\sqrt{2\pi}}\int_{-y_\epsilon}^{y_\epsilon}e^{-\frac{y^2}{2}}dy=1-\epsilon.
\end{equation}
Then we have the probability
\begin{align}
&P(-y_\epsilon<Y<y_\epsilon)\notag\\
&=P(-y_\epsilon<\frac{\mean{X}-m}{s/\sqrt{n}}<y_\epsilon)\notag\\
&=P(\mean{X}-\frac{sy_\epsilon}{\sqrt{n}}<m<\mean{X}+\frac{sy_\epsilon}{\sqrt{n}}).\label{E:est2}
\end{align}
Therefore, combining Eq.~\eqref{E:est} and Eq.~\eqref{E:est2}, we have
\begin{equation}
P(\mean{X}-\frac{sy_\epsilon}{\sqrt{n}}<m<\mean{X}+\frac{sy_\epsilon}{\sqrt{n}})=1-\epsilon.
\end{equation}
So, we can say that $(\mean{x}-\frac{sy_\epsilon}{\sqrt{n}}, \mean{x}+\frac{sy_\epsilon}{\sqrt{n}})$ is an approximate confidence interval for the distribution mean $m$ with confidence level $1-\epsilon$.

Now we can decide the photon count distribution as sub-Poissonian, or super-Poissonian, with confidence level $1-\epsilon$, if the estimated variance, $s^2\leq\mean{x}-\frac{sy_\epsilon}{\sqrt{n}}$, or $s^2\geq\mean{x}+\frac{sy_\epsilon}{\sqrt{n}}$, i.e., $\frac{\mean{x}}{s^2}\geq1+\frac{y_\epsilon}{s\sqrt{n}}$, or $\frac{\mean{x}}{s^2}\leq1-\frac{y_\epsilon}{s\sqrt{n}}$ respectively, where $y_\epsilon$ is given by Eq.~\eqref{E:est}. That is, for some given $\epsilon$ and $y_\epsilon$ given by Eq.~\eqref{E:est},
\begin{equation}
\label{E:int_est}
\begin{aligned}
\frac{\mean{x}}{s^2}\geq1+\frac{y_\epsilon}{s\sqrt{n}}\implies&\text{sub-Poissonian distribution},\\
\frac{\mean{x}}{s^2}\leq1-\frac{y_\epsilon}{s\sqrt{n}}\implies&\text{super-Poissonian distribution},
\end{aligned}
\end{equation}
with confidence level $1-\epsilon$.

The performance of this process depends on how the sample mean and sample variance converge to the distribution mean and the distribution variance as $n$ increases. The variance of the sample mean and the sample variance are given by $var[\mean{X}]=\frac{\sigma^2}{n}$ and $var[S^2]=\frac{1}{n}\left(\mu_4-\frac{n-3}{n-1}\sigma^4\right)$, respectively, where $\mu_4$ is the fourth order central moment defined as $\mu_4:=\mean{(X-\mean{X})^4}$~\cite{ross2009introduction, casella2002statistical, mood1974}. As $\frac{var[\mean{X}]}{var[S^2]}\approx\mathcal{O}(\sigma^2)$, relative convergence of the mean and the variance depends on the distribution variance. Thus, if $\sigma^2\gg1$, this process may fail to identify the nature of the distribution.

To overcome the above limitation, we suggest an alternative approach to compare the mean with the interval of the variance. Here, we find an interval for the distribution variance up to a given confidence and compare the mean with it. Suppose the sample contains $n$ data points for the number of photon counts within a fixed time. Let $\mean{x}$ and $S^2$ be the sample mean and sample variance, respectively. Then $\mean{x}$ is an unbiased estimate for the distribution mean~\cite{ross2009introduction, casella2002statistical}. It is also consistent if the variance is known. I $\sigma$ is the variance of the distribution, the probability distribution of the variance is given by $\frac{nS^2}{\sigma^2}$, which follows $\chi^2$ distribution. Then for the confidence level $1-\epsilon$, we have two numbers $\chi^2_{n-1,1-\epsilon/2}$ and $\chi^2_{n-1,\epsilon/2}$ such that
\begin{align}
&P(\chi^2_{n-1,1-\epsilon/2}<\chi^2<\chi^2_{n-1,\epsilon/2})=1-\epsilon\label{E:v_est}\\
\iff&P(\chi^2_{n-1,1-\epsilon/2}<\frac{nS^2}{\sigma^2}<\chi^2_{n-1,\epsilon/2})=1-\epsilon\notag\\
\iff&P\left(\frac{nS^2}{\chi^2_{n-1,\epsilon/2}}<\sigma^2<\frac{nS^2}{\chi^2_{n-1,1-\epsilon/2}}\right)=1-\epsilon.\label{E:v_est1}
\end{align}

Therefore, $\left(\frac{nS^2}{\chi^2_{n-1,\epsilon/2}}, \frac{nS^2}{\chi^2_{n-1,1-\epsilon/2}}\right)$ is the confidence interval for the distribution variance up to the confidence interval $1-\epsilon$. Thus, for some given $\epsilon$, we can find $\chi^2_{n-1,1-\epsilon/2}, \chi^2_{n-1,\epsilon/2}$ by Eq.~\eqref{E:v_est} and conclude that
\begin{equation}
\begin{aligned}
\mean{x}>\frac{nS^2}{\chi^2_{n-1,1-\epsilon/2}}\implies&\text{sub-Poissonian distribution},\\
\mean{x}<\frac{nS^2}{\chi^2_{n-1,\epsilon/2}}\implies&\text{super-Poissonian distribution},
\end{aligned}
\end{equation}
with confidence level $1-\epsilon$.

Alternatively, one can perform hypothesis testing to determine whether the light is sub-Poissonian, Poissonian, or super-Poissonian.

\subsection{Hypothesis Testing as an Alternative to Interval Estimation}

To determine the quantumness, one has to perform two consecutive hypothesis tests. For the first test,
\begin{equation}
\label{E:test1}
H_0:\ m=s^2,\ H_a:\ m<s^2,
\end{equation}
and for the second test,
\begin{equation}
\label{E:test2}
H_0:\ m=s^2,\ H_a:\ m>s^2,
\end{equation}
where $m$ is the mean of the actual distribution and $s^2$ is given by Eq.~\eqref{E:est_var}. For both tests, the test statistic would be the dispersion index~\cite{Fisher1992, Alves2022} $\chi^2=\frac{(n-1)s^2}{\mean{x}}$. For the first test,\\
\emph{case 1:} if $\chi^2\geq\chi^2_{n-1,1-\epsilon}$, where $\chi^2_{n-1,1-\epsilon}$ is the critical Chi-square value with $n-1$ degrees of freedom at confidence level $1-\epsilon$, $H_0$ is rejected. That implies the underlying photon distribution is sub-Poissonian.\\
\emph{case 2:} if $\chi^2<\chi^2_{n-1,1-\epsilon}$, we cannot reject $H_0$. Also, we cannot accept $H_0$ due to lack of evidence. In that case, we have to perform a second hypothesis test with the same statistic. After the second test, if $\chi^2\leq\chi^2_{n-1,\epsilon}$, $H_0$ is rejected again, and we can say that the underlying distribution is super-Poissonian.

\subsection{\label{sec:goodnes}Second Phase: Goodness-of-fit Test}

Since the equality of mean and variance may not imply Poisson distribution, to ensure the Poisson distribution which is an indicator of a coherent source of light, we suggest a second test which is a goodness-of-fit test, that is, a hypothesis testing to decide whether a sample data comes from a Poisson distribution or not.

For testing the quantumness in the generated random number, one may use the $\chi^2$ goodness-of-fit test~\cite{hogg2019introduction,chisq} on the data generated by QRNGs as follows. Consider null and alternative hypotheses as\\
$H_0$: The data follow the distribution~\eqref{E:Psn}.\\
$H_\text{a}$: The data do not follow the distribution~\eqref{E:Psn}.\\
The test statistic is defined as 
\begin{equation}
\label{E:test_stat}
\chi^2=\sum_{n=1}^\infty\frac{(O_n-E_n)^2}{E_n},
\end{equation}
where $O_n$ denotes the frequency of $n$ number of photon in the data, $E_n$ denotes the frequency given by Poisson distribution~\eqref{E:Psn}. This test statistic follows $\chi^2$ distribution with $k=N-1$ degrees of freedom if all the parameters are known, and its density function is given by,
\begin{equation}
\label{E:chi}
\begin{aligned}
&f_{\chi^2}(x)=\frac{x^{\frac{k}{2}-1}e^{-\frac{x}{2}}}{2^\frac{k}{2}\Gamma(\frac{k}{2})},\quad x\geq0,\\
\text{where }&\Gamma(z)=\int_0^\infty t^{z-1}e^{-t}dt
\end{aligned}
\end{equation}
is the well-known Gamma function~\cite{hogg2019introduction}. The cumulative distribution function in this case is given by,
\begin{equation}
\label{E:Chi}
F_{\chi^2}(z)=\int_0^zf_{\chi^2}(x)dx.
\end{equation}

\begin{figure}
\centering
\includegraphics[width = \columnwidth]{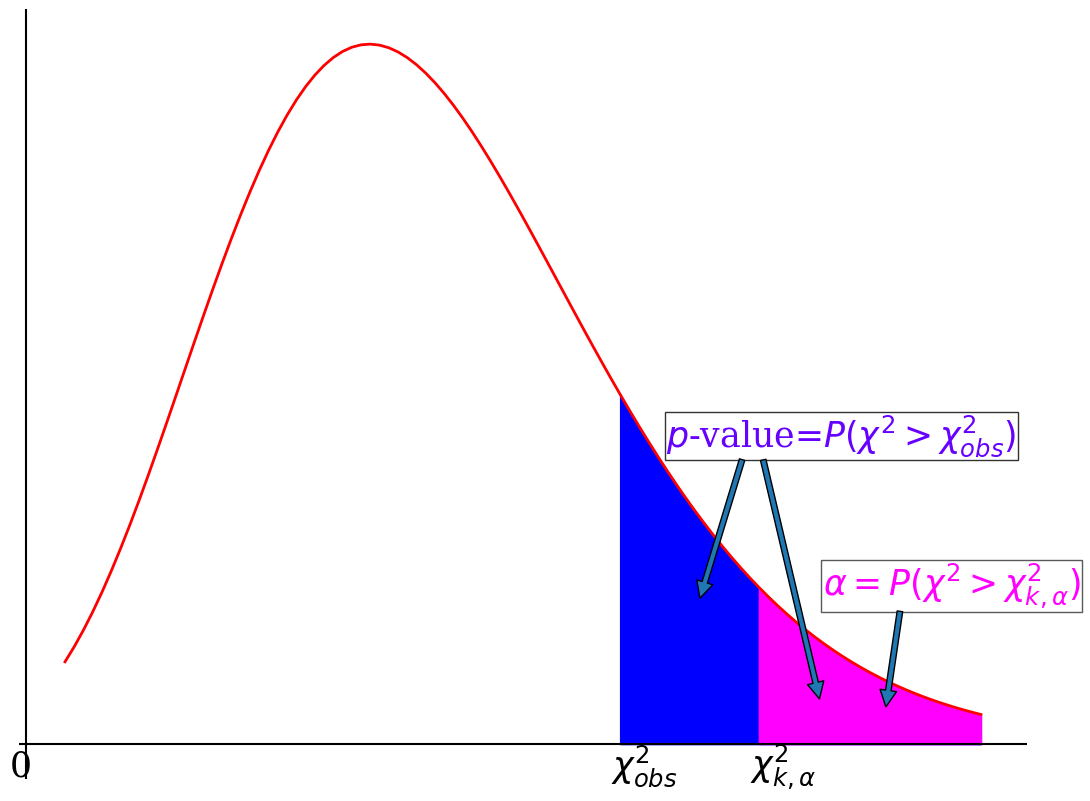}
\caption{Probability density function~\eqref{E:chi} for $\chi^2$ distribution generated using Python packages scipy and matplotlib. The critical region, $\chi^2>\chi^2_{k,\alpha}$, for the $\chi^2$ test is denoted by light shed (pink color); and the region, $\chi^2>\chi^2_{obs}$, corresponding to the $p$-value is denoted by dark as well as light sheds (blue as well as pink colors). Since here $\chi^2_{k,\alpha}>\chi^2_{obs}$, the $p$-value becomes larger than $\alpha$ and the region corresponding to the $p$-value contains the critical region. Therefore, in this case, the null hypothesis is accepted.}
\label{fig:chisq}
\end{figure}

Figure~\ref{fig:chisq} is a plot of the probability density function of a $\chi^2$ distribution. The critical region ($\chi^2>\chi^2_{k,\alpha}$) for the $\chi^2$ test is denoted by light shed (pink color), and the region corresponding to the $p$-value ($\chi^2>\chi^2_{obs}$) is denoted by dark as well as light sheds (blue as well as pink colors)‌. The area of the critical region is given by
\begin{equation}
P(\chi^2>\chi^2_{k,\alpha})=1-F_{\chi^2}(\chi^2_{k,\alpha})=\alpha.
\end{equation}
$\chi^2$-test being a one-tailed test, $p$-value for this test is given by
\begin{equation}
P(\chi^2>\chi^2_{obs})=1-F_{\chi^2}(\chi^2_{obs}),
\end{equation}
where $\chi^2_{obs}$ is the observed value of the test statistic.

If the typical values of the parameters $\mu, T$ in Eq.~\eqref{E:Psn} are known, we have $k=N-1$. However, if they are unknown, in that case, $k=N-2$. Therefore, $H_0$ is rejected if $\chi^2_{obs}>\chi^2_{k,\alpha}$, where $\chi^2_{k,\alpha}$ is the value given by $F_{\chi^2}(\chi^2_{k,\alpha})=1-\alpha$. In other words, $H_0$ is rejected if the $p$-value is less than $\alpha$.

Thus, our two-fold statistical method ensures the distribution of the photon statistics to be sub-Poissonian, Poissonian, or super-Poissonian up to a certain confidence level. As in most cases, the equality of the mean and the variance denotes a Poisson distribution; for practical purposes, the last $\chi^2$ test may not be needed. We mention this test to ensure the completeness of the discussion.

Note that if synthetic data is generated via simulation according to the actual photon statistics distribution, our two-fold method might inaccurately classify this data as coming from a sub-Poissonian quantum source. We do not assert that our two-fold method can classify any off-the-shelf black-box random number generator. Rather, we claim that, given the internal parameters and intermediate distributions from a real device, our two-fold method provides a more rigorous validation of whether the optical source operates in the classical or quantum regime compared to merely comparing the exact equality of mean and variance, thus confirming the quantum nature of the generated random numbers.

\begin{table*}[htpb]
\centering
\begin{tabular}{|c|c|c|c|c|}
\hline
\multirow{4}{*}{\textbf{Methods}}&\multicolumn{4}{c|}{\textbf{Distributions with mean $m$ = variance $\sigma^2$}}\\
\cline{2-5}
&\textbf{Geometric}&\multicolumn{2}{c|}{\textbf{Poisson}}&\textbf{Normal}\\
\cline{2-5}
&\textbf{Dataset 0 }&\textbf{Dataset 1}&\textbf{Dataset 2}&\textbf{Dataset 3}\\
&($p=0.5$)&($m=0.5$)&($m=10$)&($m=\sigma^2=0.5$)\\
\hline
Direct comparison&0.5\%&4.2\%&0.0\%&3.3\%\\
\hline
Comparison and GoF&100\%&100\%&100\%&100\%\\
\hline
Dispersion&81.3\%&98.1\%&98.2\%&90.8\%\\
\hline
Dispersion and GoF&100\%&99.4\%&99.5\%&100\%\\
\hline
Our Method (Phase-I)&85.3\%&98.9\%&99.1\%&94.0\%\\
\hline
Our Method (Phase-II)&100\%&99.4\%&99.4\%&100\%\\
\hline
\end{tabular}
\caption{Comparison for the success rates of different methods to identify the nature of the underlying distributions, all having equal mean and variance. The methods without a goodness-of-fit (GoF) test are capable only of concluding whether the mean and the variance are sufficiently close. Given that the mean and the variance are found to be close enough, the GoF test (Phase-II) confirms whether the underlying distribution is Poisson or not.  The dispersion method~\cite{Fisher1992, Alves2022} based on the dispersion index identifies the distribution with high accuracy. However, our proposed method outperforms these methods. Note that a success rate of 0\% indicates that the corresponding method fails to correctly predict the underlying distribution for all samples in the particular dataset. Similarly, a success rate of 100\% indicates the correct prediction by the corresponding method for all samples in the particular dataset.}
\label{tab:exp}
\end{table*}

\subsection{Experimental Results}

We have experimented to validate our method with synthetic datasets generated by Python~3~\cite{10.5555/1593511}. We have generated one dataset from geometric distribution~\eqref{E:geo} with $p=0.5$, and one dataset from normal distribution with $m=\sigma^2=0.5$. Also, we have considered two different means, 0.5 and 10, to generate two sets of data for the Poisson distribution. We have used the numpy library~\cite{harris2020NumPy} to generate 1000 samples, each of size 100K, for each of the datasets. Table~\ref{tab:exp} shows the performance comparison between the existing methods and our proposed method with $99\%$ confidence. The methods without a goodness-of-fit (GoF) test are capable only of concluding whether the mean and the variance are sufficiently close. If they are found close in $s$ out of $S$ (= 1000 in our case) samples, the success rate of Phase-I is given by $\frac{s}{S}$. The GoF test (Phase-II) is run on these $s$ samples to confirm whether the underlying distribution is Poisson or not. If it correctly predicts Poisson or non-Poisson in $t$ out of the above $s$ samples, the success rate of Phase-II is given by $\frac{t}{s}$.

For example, as shown in Table~\ref{tab:exp}, the direct comparison method for Dataset~0, generated from a geometric distribution, yields $0.5\%$ success rate. It means that in only $0.5\%$ of the 1000 samples, the direct comparison method concludes that the mean and the variance are close enough to decide that they are equal with $99\%$ confidence. Among these $1000\times0.5\%=5$ particular samples, the goodness-of-fit test for Poisson decides for $100\%$ of them (i.e., for all the above 5 samples) that the data is not generated from a Poisson distribution.

On the other hand, for Dataset~2 generated from the Poisson distribution, our interval estimation process (Phase-I) decides that the mean and the variance are close enough in $99.1\%$ of the samples with $99\%$ confidence. Among these $1000\times99.1\%=991$ particular samples, the goodness-of-fit test for Poisson decides for $99.4\%$ of them (i.e., for 985 of the above 991 samples) that the data is indeed generated from a Poisson distribution.

\section{\label{sec:muTd}Detailed Photon Counting Statistics of Coherent Pulse-based QRNG}

The statistics discussed in Eq.~\eqref{E:ideal} of Section~\ref{sec:coh_pulse} consider that all of the $n$ photons present within time $T$ would be detected in the same time-bin, which is a crude assumption for $n>1$. In this section, we perform a more fine-grained analysis of the first photon arrival time for coherent pulse-based QRNG devices. This analysis helps us to investigate the quality of QRNG in terms of several parameters, as elaborated in Section~\ref{sec:constraint}.

\subsection{Time-bin Statistics for Poissonian Photon with Perfect Devices}

Currently available photon arrival time-based QRNGs~\cite{quRNG,TROPOS} use coherent sources, which generate photons that follow Poissonian statistics. Here, a random number is generated based on the detection time of the first photon in each time cycle $T$. The probability of detecting the first photon in $i$-th bin, when $n$ photons are present within $T$ time, would be $P(i;1st|n) =\left(1-\frac{i-1}{N}\right)^n-\left(1-\frac{i}{N}\right)^n$.  This was observed by Nie \emph{et al.}~\cite[Eq.~(2)]{10.1063/1.4863224}. However, we do not know the exact number of photons present within the period $T$. Therefore, the probability of detecting the first photon in the $i$-th bin is given by
\begin{widetext}
\begin{align}
P(i;1st)&=\sum_{n\geq0}P(i;1st|n)P(n,T)\notag\\
&=P(i;1st|0)P(0,T)+\sum_{n>0}P(i;1st|n)P(n,T)\notag\\
&=0+\sum_{n>0}\frac{e^{-\mu T}(\mu T)^n}{n!}\left[\left(1-\frac{i-1}{N}\right)^n-\left(1-\frac{i}{N}\right)^n\right]\notag\\
&=e^{-\mu T}\sum_{n>0}\frac{\left[\mu T\left(1-\frac{i-1}{N}\right)\right]^n}{n!}-e^{-\mu T}\sum_{n>0}\frac{\left[\mu T\left(1-\frac{i}{N}\right)\right]^n}{n!}\notag\\
&=e^{-(i-1)\frac{\mu T}{N}}-e^{-i\frac{\mu T}{N}}. \label{E:act}
\end{align}
\end{widetext}

\subsection{\label{Ssec:err_QRNG}Time-bin Statistics for Poissonian Photon with Imperfect Devices}

Til now, we have considered all of the used optical devices as ideal. But in practice, any error in the source and in the attenuators affects photon power and changes the value of the mean photon number per unit time. Again, if the detector is not $100\%$ efficient, the first photon would be detected in the $i$-th time-bin, if at least one photon has been detected in $i$-th bin, and either (A) no photon has been reached at SPD before the $i$-th bin, or (B) $k$ number of photons have reached before the $i$-th bin, but none of them has been detected. Therefore, the probability of detecting the first photon in the $i$-th bin, when $n$ photons are present within $T$ time, is given by
\begin{widetext}
\begin{align}
P_d(i;1st|n)&=\frac{\sum\limits_{k=0}^{n-1}P(k,\frac{(i-1)T}{N})(1-d)^k\sum\limits_{l=1}^{n-k}\{1-(1-d)^l\}P(l,\frac{T}{N})P(n-k-l,T-\frac{iT}{N})}{P(n,T)}\notag\\
&=\sum_{k=0}^{n-1}\binom{n}{k}\left(\frac{(i-1)(1-d)}{N}\right)^k\left[\left(1-\frac{i-1}{N}\right)^{n-k}-\left(1-\frac{i-1+d}{N}\right)^{n-k}\right]\notag\\
&=\left(1-(i-1)\frac{d}{N}\right)^n-\left(1-i\frac{d}{N}\right)^n, \label{E:detdist}
\end{align}
\end{widetext}
where $d\in(0,1]$ is the detection efficiency. Since the exact number of photons is unknown, the probability of detecting the first photon in the $i$-th bin becomes
\begin{align}
\label{E:det}
P_d(i;1st)&=e^{-(i-1)\frac{\mu Td}{N}}-e^{-i\frac{\mu Td}{N}}=(e^\frac{\mu Td}{N}-1)e^{-i\frac{\mu Td}{N}}.
\end{align}
The remaining probability $1-\sum_{i=1}^NP_d(i;1st)=e^{-\mu Td}$ is the probability of detecting no photon in a complete reference cycle. In that case, this particular reference cycle has no contribution to the random number. Therefore, the probability mass function of the distribution corresponding to the detection of the first photon in the $i$-th bin is given by
\begin{equation}
\label{E:distIII}
f(i)=\frac{e^\frac{\mu Td}{N}-1}{1-e^{-\mu Td}}e^{-i\frac{\mu Td}{N}},\;i=1,2,\dots,N.
\end{equation}

\begin{figure}
\centering
\includegraphics[width=\columnwidth]{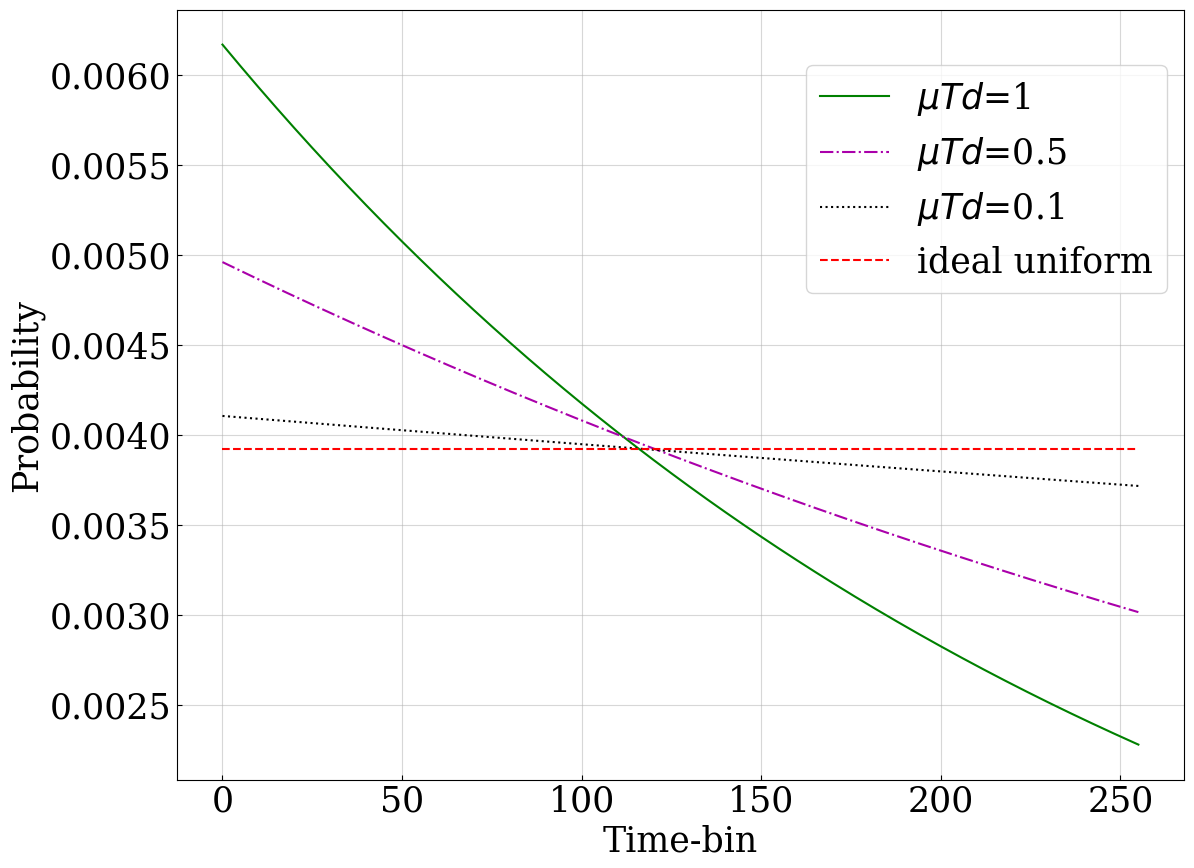}
\caption{Plots for the probabilities of detecting the first photon in $256$ time-bins given by Eq.~\eqref{E:distIII} with Python matplotlib library~\cite{hunter2007matplotlib}. Probabilities for different values of $\mu Td$ as well as uniform probability distribution have been plotted. $\mu$ is the expected number of photons in unit time, $T$ is the time cycle in the external reference time frame, and $d$ is the detection efficiency. If all the optical components are ideal, the distribution should be given by the red dashed line.}
\label{fig:Prob_plot}
\end{figure}

Note that $f(1)$ is the probability $P_1$ in Ref.~\cite[Eq.~(4)]{10.1063/1.4863224}. As their aim was to find the minimum entropy for deciding the amount of post-processing, they only derived an upper bound to $P_1$. However, we aim to reduce the amount of post-processing. Therefore, we have derived the complete probability mass function to see how the device parameters, $\mu, T$ and $d$ affect the distribution. We also discuss the effect of all the parameters separately in Section~\ref{sec:constraint}. Figure~\ref{fig:Prob_plot} shows how $\mu Td$ affects the probability~\eqref{E:distIII}. Here we consider $N=256$. From this plot, it is clear that the uniformity increases as $\mu Td$ decreases.

Although this uniformity can be increased by post-processing the raw data, as mentioned above, this reduces the random number generation rate and increases the possibility of a side-channel attack. Since here we want to minimize the post-processing, the main goal in designing a QRNG is
\begin{equation}
\label{E:goal}
\begin{aligned}
&\underset{\mu, T, d}{minimize}\ g(\mu, T, d):=\mu Td,\\
&subject\ to\ \mu>0,T>0,d>0.
\end{aligned}
\end{equation}
However, this minimization is bounded by some constraints related to the generation rate, the cost and the timing error in registering the photon detection time. These constraints are discussed in Section~\ref{sec:constraint}.

Next, let us consider QRNG in Fig.~\ref{fig:time_diag}(II). Here, the probability, that the waiting time ($t_w$) between the arrival of two consecutive photons lies in the $i$-th bin is given by $P((i-1)t_l<t_w<it_l)=P(t_w>(i-1)t_l)-P(t_w>it_l)=e^{-(i-1)\mu t_l}-e^{-i\mu t_l}$, where $t_l$ is the length of each time-bin. If the detection efficiency is given by $d$, this probability is given by
\begin{equation}
\label{E:detII}
P_d((i-1)t_l<t_w<it_l)=e^{-(i-1)\mu t_ld}-e^{-i\mu t_ld}.
\end{equation}
Since in the probability distribution of Eq.~\eqref{E:det}, $\frac{T}{N}$ is the length of the time-bins, Eq.~\eqref{E:detII} is identical with Eq.~\eqref{E:det}. Therefore, the probability mass function corresponding to the probability Eq.~\eqref{E:detII} is also given by Eq.~\eqref{E:distIII} with $T=Nt_i$.

Therefore, these two types of QRNGs perform similarly in practical scenarios. However, if we can ensure that the source is an actual single-photon source, then the perfect random number can be generated using an external time reference with probability distribution $P_d(i)=\frac{d}{N}$ (Ref. Eq.~\eqref{E:ideal}). Since the distributions given by Eq.~\eqref{E:distIII} are evaluated from the quantum properties of photons, this can be used in the goodness-of-fit test mentioned in Subsection~\ref{sec:goodnes} to test the quantumness in the raw data generated by a QRNG.

\begin{definition}\label{def:randomness}
We define a probability distribution $P(\cdot)$, with support $S$, as \emph{$\epsilon$-random} if and only if
\begin{equation}
\begin{aligned}
\max_{V\subseteq S}\left|\sum_{i\in V}\left(P(i)-\frac{1}{|S|}\right)\right|
=\frac{1}{2}\sum_{i\in S}\left|P(i)-\frac{1}{|S|}\right|&=\epsilon.
\end{aligned}
\end{equation}
\end{definition}

\begin{figure}
\centering
\includegraphics[width=\columnwidth]{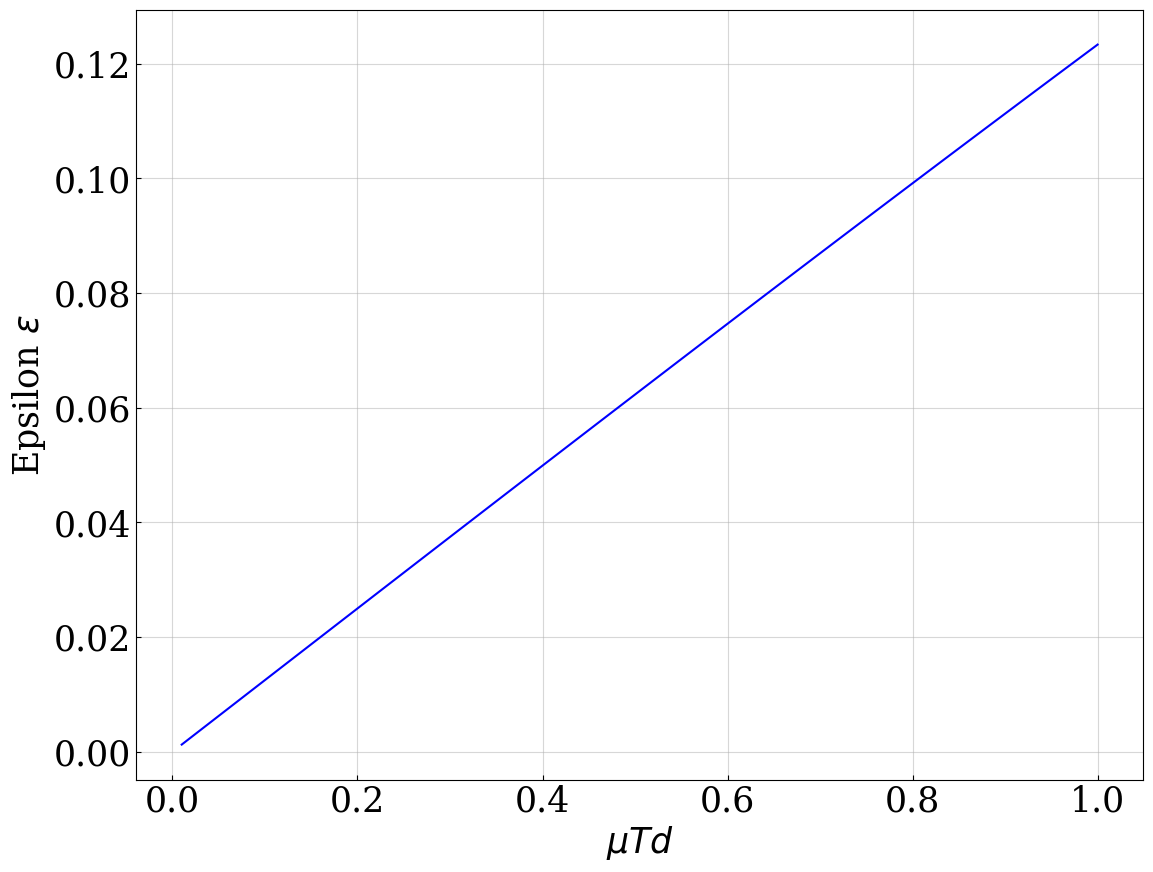}
\caption{Plot representing the relation between $\mu Td$ and $\epsilon$ such that the probability distribution is $\epsilon$-random given by Eq.~\eqref{E:rand} with Python matplotlib library~\cite{hunter2007matplotlib}. Randomness reduces as $\mu Td$ increases. This product can be reduced with the proper choice of external reference cycle $T$, and proper tuning of the source, the attenuators and the detector.}
\label{fig:randomness}
\end{figure}

The distribution~\eqref{E:distIII} becomes $\epsilon$-random if and only if
\begin{equation}
\frac{1}{2}\sum_{i\in S}\left|f(i)-\frac{1}{N}\right|=\epsilon. \label{E:rand}
\end{equation}
Figure~\ref{fig:randomness} is a visual representation of the Eq.~\eqref{E:rand} with $N=256$. It clearly shows that the value of $\epsilon$ increases with $\mu Td$. That means, to avoid post-processing while generating random numbers with $\epsilon$-randomness for a given $\epsilon$, one needs to carefully choose the external reference cycle $T$, and configure the source, the attenuators and the detector so that relation~\eqref{E:rand} is satisfied.

We have performed the NIST, Dieharder, AIS-31 and ENT tests on the data generated by simulating the probability distribution given by Eq.~\eqref{E:distIII}. These test results are provided in Section~\ref{sec:NIST}.

\section{\label{sec:constraint}Effect of $\mu,T,d$ on random number generation rate and cost}

As discussed in Eq.~\eqref{E:goal} of Subsection~\ref{Ssec:err_QRNG}, we have to minimize $\mu Td$ to achieve maximum randomness from the quantum source increases. In that discussion, we have considered the product $\mu Td$ as a single quantity. In this section, we separately discuss the effect of
\begin{itemize}
\item the expected photon number in unit time, $\mu$,
\item the external reference cycle, $T$, and
\item the detection efficiency, $d$
\end{itemize}
on the rate of random number generation, $r$ as well as on the cost of the QRNG.

The randomness of an RNG is characterized by the \emph{entropy} of the underlying probability distribution followed by the RNG. Let $H(Q)$ be the entropy generated by a quantum source. After postprocessing it becomes
\begin{equation}
\label{E:randomness}
H(Q)\xrightarrow{postprocessing}H(E),
\end{equation}
where $H(E)$ is the final entropy generated by the QRNG. As deterministic classical post-processing cannot improve the entropy, in general $H(E)\leq H(Q)$.

\begin{definition}
An RNG is said to be \emph{ideal} if $H(E)=H(U)$, where $H(U)$ is the entropy of uniform randomness. A QRNG is said to be \emph{ideal} if $H(U)=H(E)=H(Q)$.
\end{definition}

However, in the practical scenario, imperfections within used hardware components make an ideal QRNG impossible. Therefore, the main goal in designing a QRNG is
\begin{equation}
\label{E:goal_ent}
\textbf{Goal: }H(Q)\approx H(U).
\end{equation}

The Shannon entropy~\cite{6773024,6773067} of a uniform distribution over $n$ points is given by

\begin{equation}
\label{E:ent_uni}
H(U)=-\sum_{i=1}^n\frac{1}{n}\log\frac{1}{n}=\log n.
\end{equation}
On the other hand, the entropy of the probability distribution~\eqref{E:distIII} is given by
\begin{widetext}
\begin{equation}
\label{E:act_ent}
H(Q) = \frac{1-(N+1)e^{-\mu Td}+Ne^{-(N+1)\frac{\mu Td}{N}}}{(1-e^{-\mu Td})(1-e^{-\frac{\mu Td}{N}})}\Log{e^\frac{\mu Td}{N}}-\Log{\frac{e^\frac{\mu Td}{N}-1}{1-e^{-\mu Td}}}.
\end{equation}
\end{widetext}
A detailed derivation of this entropy along with the definition of Shannon entropy and \emph{minimum entropy} has been added in~\ref{sec:entropy}. Clearly, $H(Q)\neq\log N=H(U)$. So, we have to apply post-processing as mentioned in Eq.~\eqref{E:randomness} to achieve $H(E)=H(U)$ for some uniform distribution $U$ whose support contains less than $N$ points. This post-processing, also known as randomness extractor, is a data compression procedure using universal hashing~\cite{PhysRevA.87.062327}. The minimum entropy $H_{min}$ of the raw data decides the amount of this data compression. If $H_{min}\geq\frac{m}{n}$ for some positive integers $m,n$, an $n$-bit raw data needs to be compressed to an $m$-bit data to make it uniform random~\cite{10.1063/1.4863224}. As $f(i)$ in Eq.~\eqref{E:distIII} is a decreasing function, the minimum entropy of the photon arrival-time-based QRNG discussed above would be
\begin{align}
H_{min}&=-\log f(1)\notag\\
&=-\log \frac{1-e^{-\frac{\mu Td}{N}}}{1-e^{-\mu Td}}\notag\\
&=-\log \left(1-e^{-\frac{\mu Td}{N}}\right)+\log \left(1-e^{-\mu Td}\right).\label{E:min_ent}
\end{align}
Nie \emph{et al.}~\cite[Eq.~(5)]{10.1063/1.4863224} gave a lower bound of this $H_{min}$. Since our aim is to reduce the amount of post-processing, which depends on $\mu Td$, we want to investigate $\mu, T$ and $d$ separately.

\begin{figure}
\centering
\includegraphics[width=\columnwidth]{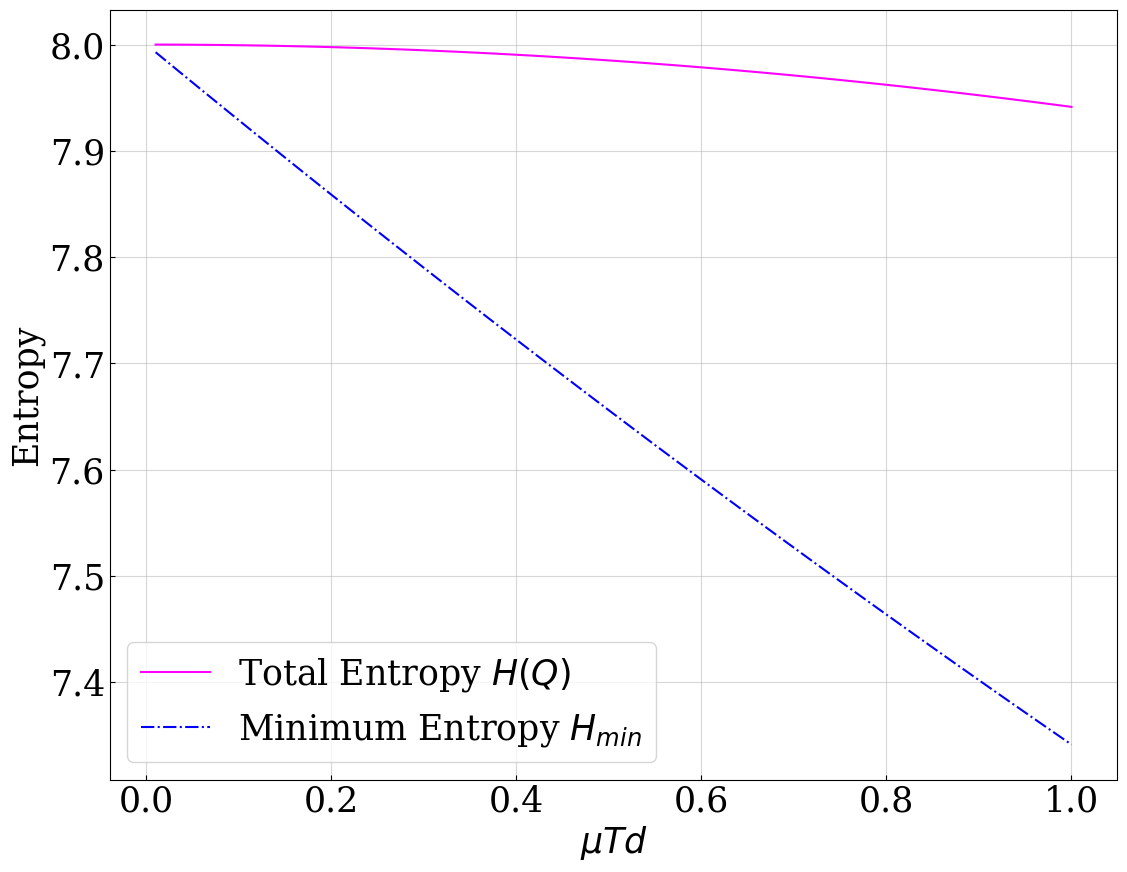}
\caption{Total entropy~\eqref{E:act_ent} and minimum entropy~\eqref{E:min_ent} of the probability distribution~\eqref{E:distIII} with Python matplotlib library~\cite{hunter2007matplotlib}. When the product of the expected number of photons, $\mu T$, within an external reference cycle $T$, and the detection efficiency $d$ is very small, $H(Q)$ and $H_{min}$ are very close to the entropy $H(U)=8$ of the uniform random distribution. Therefore, if $\mu Td$ is very small, the raw data generated by the QRNG would be close to uniform random number.}
\label{fig:ent_act}
\end{figure}

Fig.~\ref{fig:ent_act} shows how the total entropy~\eqref{E:act_ent} and minimum entropy~\eqref{E:min_ent} changes with $\mu Td$. Here we have chosen $N=256$. Therefore, the entropy~\eqref{E:ent_uni} of the uniform distribution would be $H(U) = \log 256=8$. From Fig.~\ref{fig:ent_act}, it can be easily seen that, if $\mu Td$ is very small, the entropy $H(Q)$ and $H_{min}$ are very close to $8$, which is the entropy of the uniform distribution. In this section, by \emph{randomness} we will refer to the randomness due to $H_{min}$.

\subsection{Effect of the Expected Photon Numbers}

Here we discuss the effect of the expected photon numbers on the cost as well as the random number generation rate of a QRNG.

\subsubsection{Effect of the Expected Photon Number on Cost}

Since $\mu$ is the expected number of photons in unit time, this can be reduced by using heavy attenuation. This will directly affect the cost. Heavy attenuation requires a large number of attenuators, increasing the production cost for the random numbers. Thus
\begin{equation}
\label{E:cost_mu}
\text{small value of $\mu\implies$ high production cost}.
\end{equation}

\subsubsection{Effect of the Expected Photon Number on Random Number Generation Rate}

A small value of $\mu$, which indicates a small number of photons, leads to a high chance of \emph{no photon detection} in a particular external reference cycle of time interval $T$. That means those intervals will have no contribution to the final random number. Therefore, we require more cycles to produce random numbers of expected length. This will reduce the speed of the RNG. From Eq.~\eqref{E:det}, the probability, $P(0)$, of no photon detection within time $T$ is given by

\begin{equation}
\label{E:nophoton}
P(0) = 1-\sum_{i=1}^NP_d(i;1st)=e^{-\mu Td},
\end{equation}
where $P_d(i;1st)$ is the probability of
detecting the first photon in the $i$-th bin.

Suppose, for some $\mu$, the expected number of external cycles to produce a random number of length $k$ bits, where each cycle contains $N$ time-bins, is given by $N_c$. Then $N_c=\mean{X}$, where $X$ is a random variable, whose probability distribution is a negative binomial distribution, given by
\begin{equation}
\label{E:bi_nom}
P(X=x)=\binom{x-1}{k'-1}\left(1-P_\mu(0)\right)^{k'}P_\mu(0)^{x-k'},
\end{equation}
for $x=k',k'+1,k'+2,\dots$, where $k'=\lceil\frac{k}{\log N}\rceil$ is the minimum number of cycles required to produce $k$-bit random number and $P_\mu(0)=e^{-\mu Td}$. Therefore,
\begin{equation}
\label{E:num_cycle}
N_c=\mean{X}=\frac{k'}{1-P(0)}=\frac{k'}{1-e^{-\mu Td}}.
\end{equation}
Thus, the rate of generating random numbers is given by
\begin{equation}
\label{E:rate}
r=\frac{k}{TN_c}\approx\frac{1-e^{-\mu Td}}{T}\log N\text{ bits per unit time}.
\end{equation}
Since the external time cycle $T$ is very small, Eq.~\eqref{E:rate} can be written as
\begin{equation}
\label{E:murate}
r\approx \mu d\log N\text{ bits per unit time}.
\end{equation}

\begin{figure}
\centering
\includegraphics[width=\columnwidth]{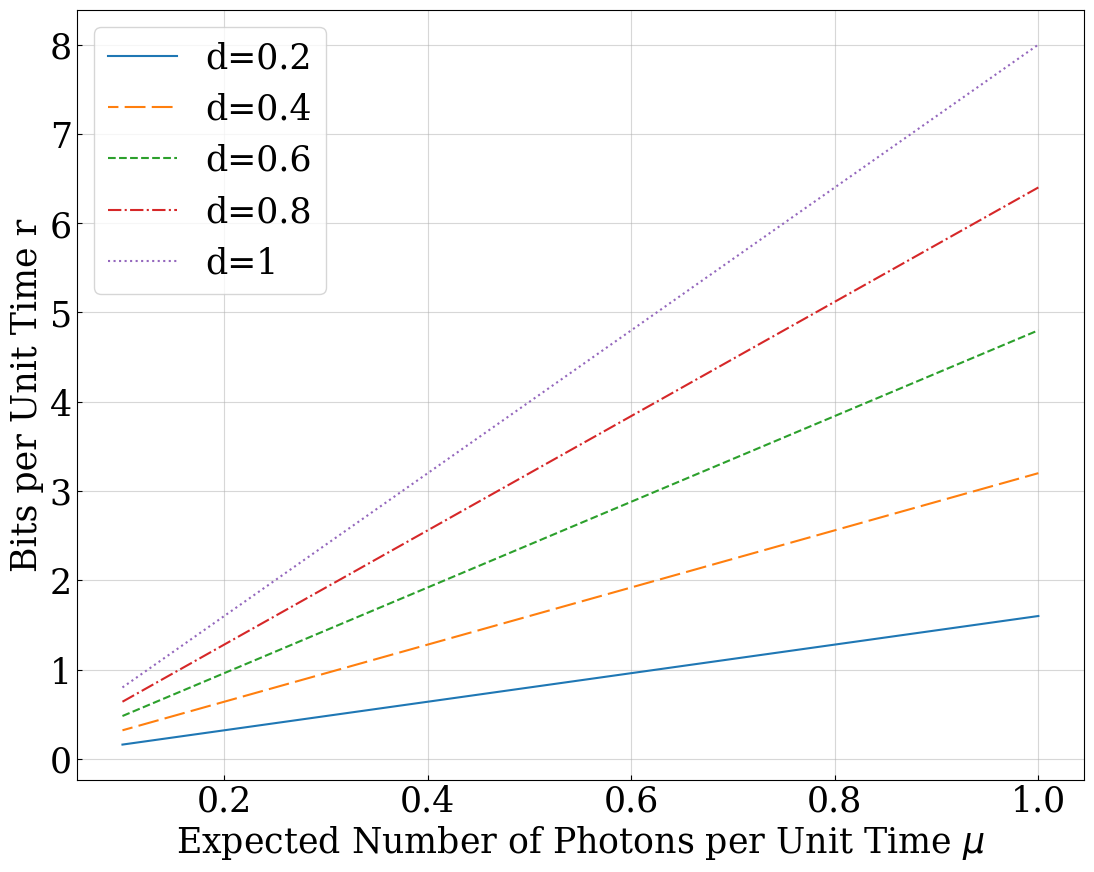}
\caption{Plot shows how random number generation rate (bit per unit time) changes with the expected photon number per unit time ($\mu$) given by Eq.~\eqref{E:murate} when $d$ is constant. Five different plots have been shown for five different values of detection efficiency ($d$). From the plot, it can be easily seen that if $\mu$ decreases, the random number generation rate of the QRNG will also decrease.}
\label{fig:murate}
\end{figure}

Clearly, if $\mu$ decreases, this rate will also decrease. The graph in Fig.~\ref{fig:murate} shows how this rate changes with $\mu$ for fixed $T$ and $d$. Therefore, we cannot reduce $\mu$ arbitrarily to get the maximum amount of randomness.

\subsection{Effect of the timing error of the device on the External Reference Cycle}

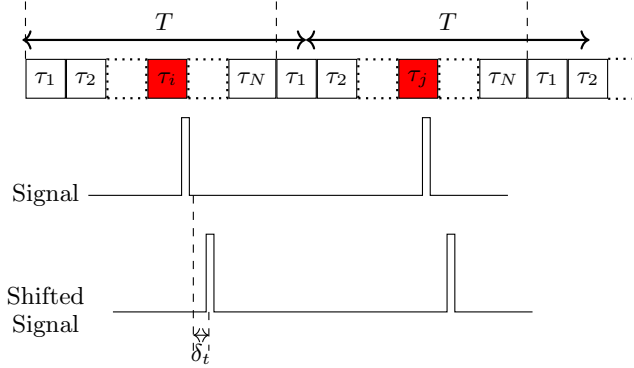
\begin{figure}
\centering
\resizebox{\columnwidth}{!}{\begin{tikzpicture}
\node[rectangle, draw, minimum size = .5cm] (1) {$\tau_1$};
\node[rectangle, draw, right = 0cm of 1, minimum size = .5cm] (2) {$\tau_2$};
\node[rectangle, thick, draw, dotted, right = 0cm of 2, minimum size = .5cm] (3) {};
\node[rectangle, fill = red, draw, right = 0cm of 3, minimum size = .5cm] (4) {$\tau_i$};
\node[rectangle, thick, draw, dotted, right = 0cm of 4, minimum size = .5cm] (5) {};
\node[rectangle, draw, right = 0cm of 5, minimum size = .5cm] (6) {$\tau_N$};
\node[rectangle, draw, right = 0cm of 6, minimum size = .5cm] (7) {$\tau_1$};
\node[rectangle, draw, right = 0cm of 7, minimum size = .5cm] (8) {$\tau_2$};
\node[rectangle, thick, draw, dotted, right = 0cm of 8, minimum size = .5cm] (9) {};
\node[rectangle, fill = red, draw, right = 0cm of 9, minimum size = .5cm] (10) {$\tau_j$};
\node[rectangle, thick, draw, dotted, right = 0cm of 10, minimum size = .5cm] (11) {};
\node[rectangle, draw, right = 0cm of 11, minimum size = .5cm] (12) {$\tau_N$};
\node[rectangle, draw, right = 0cm of 12, minimum size = .5cm] (13) {$\tau_1$};
\node[rectangle, draw, right = 0cm of 13, minimum size = .5cm] (14) {$\tau_2$};
\draw[dashed] (1.north west) -- ++(0, 0.75) (6.north east) -- ++(0, 0.75) (12.north east) -- ++(0, 0.75);
\draw[<->, thick] (-0.3, 0.5) -- ++(3.65,0) node[midway, above] {$T$};
\draw[<->, thick] (3.35, 0.5) -- ++(3.65, 0) node[midway, above] {$T$};
\draw[dotted, thick] (14.north east) -- ++(0.3,0);
\draw[dotted, thick] (14.south east) -- ++(0.3,0);
\node at (0, -1.5) (sig) {Signal};
\draw (sig.east) --++(1.2, 0) --++(0, 1) --++(0.1, 0) --++(0, -1) --++(3, 0) --++(0, 1) --++(0.1, 0) --++(0, -1) --++(1, 0);
\node[text width = 1.5cm, text centered] at (0, -3) (ssig) {Shifted Signal};
\draw (ssig.east) --++(1.2, 0) --++(0, 1) --++(0.1, 0) --++(0, -1) --++(3, 0) --++(0, 1) --++(0.1, 0) --++(0, -1) --++(1, 0);
\draw[dashed] (1.9, -1.5) --++(0, -2) (2.1, -3) --++(0, -0.5);
\draw[<->] (1.9, -3.3) -- (2.1, -3.3) node[midway, below] {$\delta_t$};
\end{tikzpicture}}
\caption{Due to the timing error $\delta_t$, the actual may be read as the shifted signal and the registered time-bin may be different from the actual time-bin of photon detection. This would add an error in the generated random number. To minimize such error, we put a lower bound on the external reference cycle $T$ using Eq.~\eqref{E:T_min}.}
\label{fig:Timing_error}
\end{figure}

If we choose the external reference cycle $T$ very small, the product $\mu Td$ would also be small. This should increase the randomness, as discussed in Eq.~\eqref{E:goal} of the Subsection~\ref{Ssec:err_QRNG}. However, a small $T$ would lead to a small length $\frac{T}{N}$ for the time bins. Now, if there are maximum $\delta_t$ errors in registering the photon-arrival-time, due to small time-bins, the time may be registered in a time-bin that is different from the actual time-bin as shown in Fig.~\ref{fig:Timing_error}. Let us consider that the actual detection happens in $i$-th bin, that is, within time interval $[(i-1)\frac{T}{N},i\frac{T}{N})$. If the actual detection happens within time interval $[(i-1)\frac{T}{N},(i-1)\frac{T}{N}+\delta_t)$, the time may be registered as $(i-1)$-th bin. Similarly, if the actual detection happens within time interval $[i\frac{T}{N}-\delta_t, i\frac{T}{N})$, the time may be registered as $(i+1)$-th bin. However, if the actual detection lies within the time interval $[(i-1)\frac{T}{N}+\delta_t,i\frac{T}{N}-\delta_t)$, there will be no error.

Now let us consider that $T$ is chosen to satisfy the relation $\frac{T}{N}\geq k\delta_t$, for some suitable $k$. Then the probability, $p_e$, that the detection happens in $i$-th bin but outside of the interval $[(i-1)\frac{T}{N}+\delta_t,i\frac{T}{N}-\delta_t)$ is bounded above as $p_e\leq \frac{2}{k}$.
Here we have assumed uniform distribution as the time $\frac{T}{Nk}$ is very small. Therefore, for a fixed error tolerance, $p_{tol}$, $k$ can be chosen as $k_{tol}=\frac{2}{p_{tol}}$. In that case, the minimum reference cycle $T_{min}$ should be
\begin{equation}
\label{E:T_min}
T_{min}=Nk_{tol}\delta_t.
\end{equation}
Therefore, we cannot make $T$ arbitrarily small to get the maximum amount of randomness.

\subsection{Effect of  the Reference Cycle on Random Number Generation Rate and Cost}

Equation~\eqref{E:murate} indicates that $T$ does not have much effect on the rate of generating random numbers. Also, since $T$ has no impact on the used optical devices, it does not affect cost.

\subsection{Effect of Detection Efficiency}

In this subsection, we discuss the effect of the detection efficiency of the detector used in the QRNG on the cost as well as the random number generation rate of a QRNG.

\subsubsection{Effect of  the Detection Efficiency on Cost}

The value of $\mu Td$ can be decreased by choosing a detector with a bad detection efficiency. It is very natural that if a detector can efficiently detect photons, its cost will also be high. Since the randomness increases if $\mu Td$ decreases, choosing small $d$ we can increase the randomness. This will also reduce the cost of the QRNG. Thus
\begin{equation}
\label{E:cost_d}
\text{small value of }d\implies\begin{cases}
\text{high randomness}\\
\text{low cost.}
\end{cases}
\end{equation}

\subsubsection{Effect of  the Detection Efficiency on Random Number Generation Rate}

\begin{figure}[htpb]
\centering
\includegraphics[width=\columnwidth]{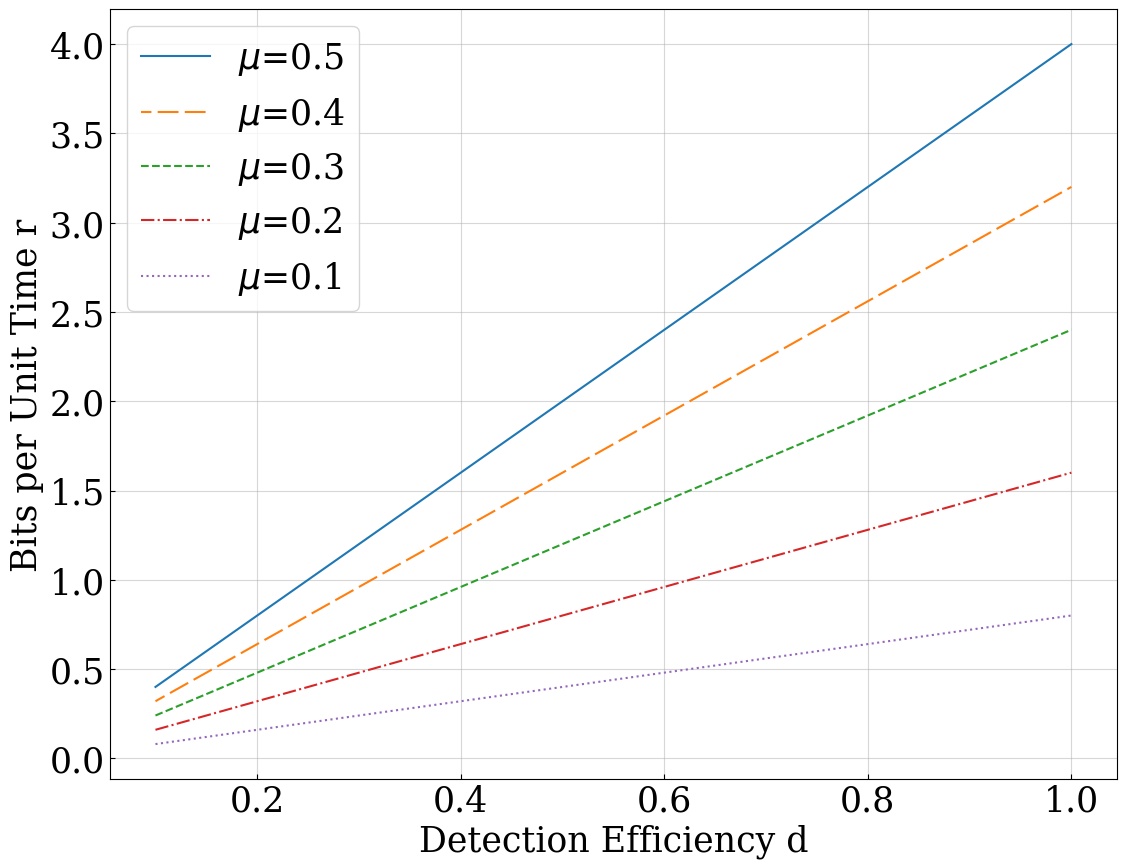}
\caption{Plot shows how random number generation rate (bit per unit time) changes with the detection efficiency ($d$) given by Eq.~\eqref{E:murate} when $\mu$ is constant. Five different plots have been shown for five different values of $\mu$, the expected photon number per unit time. From the plot, it can be easily seen that if $d$ decreases, the random number generation rate of the QRNG will also decrease.}
\label{fig:drate}
\end{figure}

Although a bad detector with a small detection efficiency, $d$, reduces production costs, this leads to a high chance of \emph{no photon detection} in a particular external reference cycle. From Eq.~\eqref{E:murate} it can be easily seen that if $T$ decreases, the rate of generating random numbers will also decrease. The graph in Fig.~\ref{fig:drate} shows how this rate changes with $d$ for fixed $T$ and $\mu$. Therefore, it is not possible to choose a detector with an arbitrarily small detection efficiency.

\section{\label{sec:eval_simult}Simultaneous Analysis of Rate, Randomness, Entropy and Cost}

\begin{figure}[htpb]
\centering
\includegraphics[width=\columnwidth]{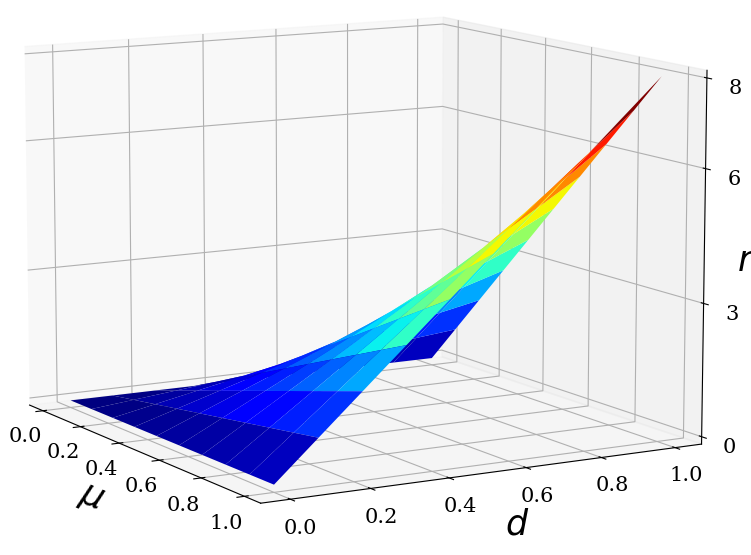}
\caption{Plot showing relation between random number generation rate, $r$, expected photon number in unit time, $\mu$, and detection efficiency, $d$ given by Eq.~\eqref{E:murate}. If the values of $\mu$ and $d$ are small, the generation rate is also small. Therefore, to generate random numbers with high rate, we have to choose both $\mu$ and $d$ large.}
\label{fig:mudrate}
\end{figure}

We can see from Eq.~\eqref{E:murate} that random number generation rate $r\approx\mu d\log N$, where $\mu$ is the expected number of photons in unit time, $d$ is the efficiency of detector and $N$ is the number of time-bins. Figure~\ref{fig:mudrate} is a surface plot of Eq.~\eqref{E:rand} to show how the random number generation rate of the QRNG changes with $\mu$ and $d$. This shows that we have to consider high values of $\mu$ and $d$ to get a high rate. On the other hand, in Eq.~\eqref{E:goal} of Subsection~\ref{Ssec:err_QRNG} we show that to get maximum randomness from the quantum source, we have to minimize the product $\mu Td$ small. Also, from Eq.~\eqref{E:cost_mu}and~\eqref{E:cost_d}, we can write the cost of the QRNG as
\begin{equation}
\label{E:cost}
\text{cost}=\frac{\alpha}{\mu}+\beta d,
\end{equation}
for some positive constants $\alpha$ and $\beta$. Therefore, this needs some optimization while choosing $\mu$ and $d$.

Considering all the discussions in the previous section, we can now write random number generation rate $r$ as a function of randomness. From Eq.~\eqref{E:distIII},~\eqref{E:rand} and~\eqref{E:murate}, we have,
\begin{widetext}
\begin{align}
2\epsilon&=\sum_{i=1}^N\left|\frac{e^\frac{\mu Td}{N}-1}{1-e^{-\mu Td}}e^{-i\frac{\mu Td}{N}}-\frac{1}{N}\right|\notag\\
&=\sum_{i\leq N/2}\left(\frac{e^\frac{\mu Td}{N}-1}{1-e^{-\mu Td}}e^{-i\frac{\mu Td}{N}}-\frac{1}{N}\right)+\sum_{i>N/2}\left(\frac{1}{N}-\frac{e^\frac{\mu Td}{N}-1}{1-e^{-\mu Td}}e^{-i\frac{\mu Td}{N}}\right)\notag\\
&=\frac{e^\frac{\mu Td}{N}-1}{1-e^{-\mu Td}}\left(\sum_{i\leq N/2}e^{-i\frac{\mu Td}{N}}-\sum_{i>N/2}e^{-i\frac{\mu Td}{N}}\right)\notag\\
&=\frac{e^\frac{\mu Td}{N}-1}{1-e^{-\mu Td}}\left(\frac{e^{-\frac{\mu Td}{N}}\left(e^{-\frac{\mu Td}{2}}-1\right)}{e^{-\frac{\mu Td}{N}}-1}-\frac{e^{-\frac{\mu Td}{N}\left(\frac{N}{2}+1\right)}\left(e^{-\frac{\mu Td}{2}}-1\right)}{e^{-\frac{\mu Td}{N}}-1}\right)\notag\\
&=\frac{\left(1-e^{-\frac{\mu Td}{2}}\right)^2}{1-e^{-\mu Td}}=\frac{1-e^{-\frac{\mu Td}{2}}}{1+e^{-\frac{\mu Td}{2}}}.\label{E:randomness_muTd}
\end{align}
\end{widetext}
Therefore we can write,
\begin{align}
&2\epsilon\left(1+e^{-\frac{\mu Td}{2}}\right)=1-e^{-\frac{\mu Td}{2}}\notag\\
\implies&e^{\frac{\mu Td}{2}}=\frac{1+2\epsilon}{1-2\epsilon}\notag\\
\implies&r=\frac{2\log N}{T}\ln{\frac{1+2\epsilon}{1-2\epsilon}}.\label{E:rate_rand}
\end{align}

\begin{figure}
\centering
\includegraphics[width=\columnwidth]{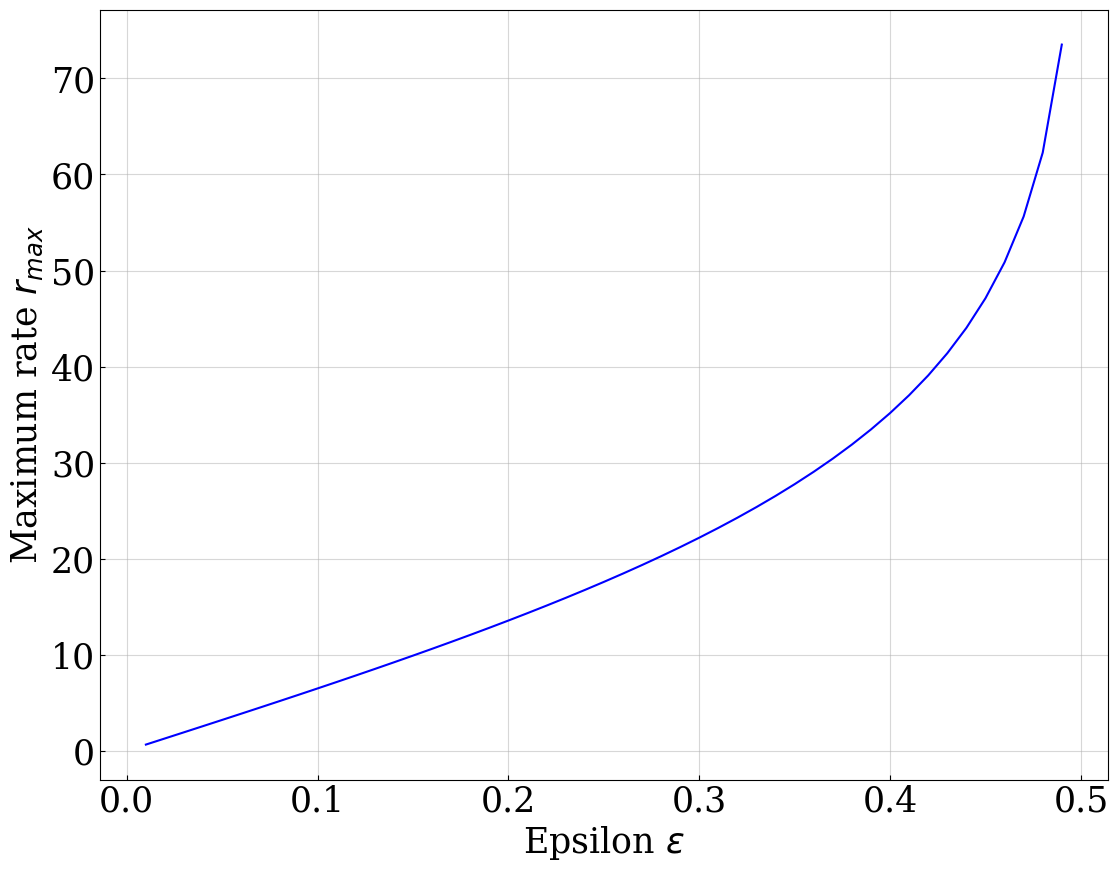}
\caption{Maximum rate is plotted against $\epsilon$ as Eq.~\eqref{E:max_rate} with $T_{min}=Nk_{tol}\delta_t=1$ unit of time. It shows that the maximum rate increases very quickly with $\epsilon$. Since from Definition~\ref{def:randomness}, randomness decreases with the increment of $\epsilon$, the maximum rate decreases as randomness increases.}
\label{fig:max_rate}
\end{figure}

Equation~\eqref{E:rate_rand} gives the random number generation rate when the generated number is $\epsilon$-random, according to Definition~\ref{def:randomness}. Therefore using Eq.~\eqref{E:T_min}, we can write the maximum rate for some given $\epsilon$ as
\begin{equation}
\label{E:max_rate}
r_{max}=\frac{2\log N}{Nk_{tol}\delta_t}\ln{\frac{1+2\epsilon}{1-2\epsilon}}\text{ bits/ unit time},
\end{equation}
where $N$ is the number of time-bins, $\delta_t$ is the maximum timing error in photon detection and $k_{tol}$ is a parameter corresponding to the tolerance of this timing error. Fig.~\ref{fig:max_rate} plots this maximum rate against $\epsilon$. It shows that the rate decreases with the increment of randomness.

\begin{figure}
\centering
\includegraphics[width=\columnwidth]{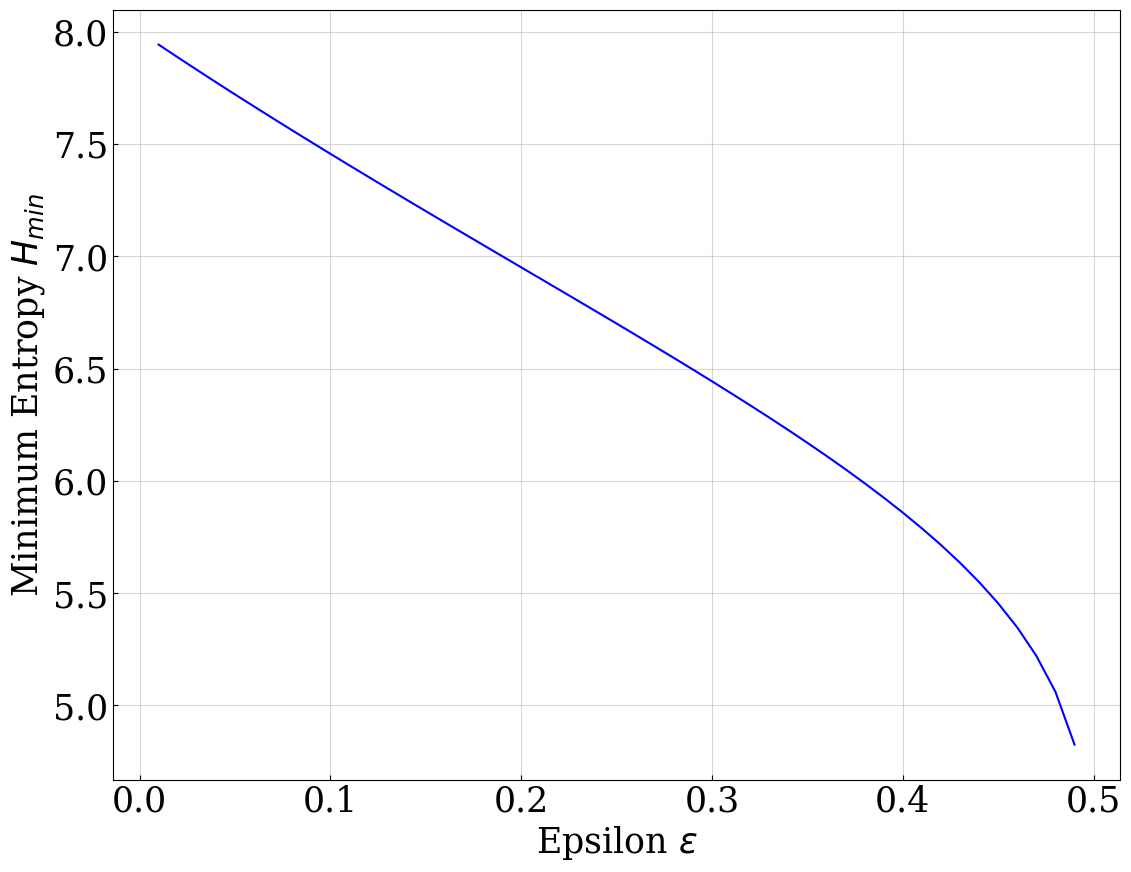}
\caption{Minimum entropy, Eq.~\eqref{E:eps_min_ent}, is plotted against $\epsilon$ for $N=256$. It shows that the minimum entropy decreases, which is a clear indication of decrement of randomness, as $\epsilon$ increases.}
\label{fig:min_ent}
\end{figure}

Also, we can write the minimum entropy~\eqref{E:min_ent} as a function of $\epsilon$ as
\begin{equation}
\label{E:eps_min_ent}
H_{min}=\log\frac{8\epsilon}{(1+2\epsilon)^2\left(1-\left(\frac{1-2\epsilon}{1+2\epsilon}\right)^\frac{2}{N}\right)}.
\end{equation}
In fig.~\ref{fig:min_ent} we plot this minimum entropy against $\epsilon$.

\begin{table}
\centering
\begin{tabular}{|c|c|c|c|}
\hline
\textbf{Parameters}&\multicolumn{3}{c|}{\textbf{Effects on}}\\
\cline{2-4}
$\downarrow$&\textbf{Randomness}&\textbf{Rate}&\textbf{Cost}\\
\hline
$\mu$&$\uparrow$&$\downarrow$&$\uparrow$\\
$T$&$\uparrow$&$-$&$-$\\
$d$&$\uparrow$&$\downarrow$&$\downarrow$\\
\hline
\end{tabular}
\caption{The effects of decreasing the expected photon count $\mu$, the external reference cycle $T$ and the detection efficiency $d$. A down-arrow, $\downarrow$, denotes the decrement of the corresponding quantity. On the other hand, an up-arrow, $\uparrow$, denotes the increment. $T$ does not affect speed or cost. If the parameters increase, the effects will be in the reverse direction.}
\label{tab:mutd}
\end{table}

\begin{figure*}[thpb]
\centering
\includegraphics[width=0.49\textwidth]{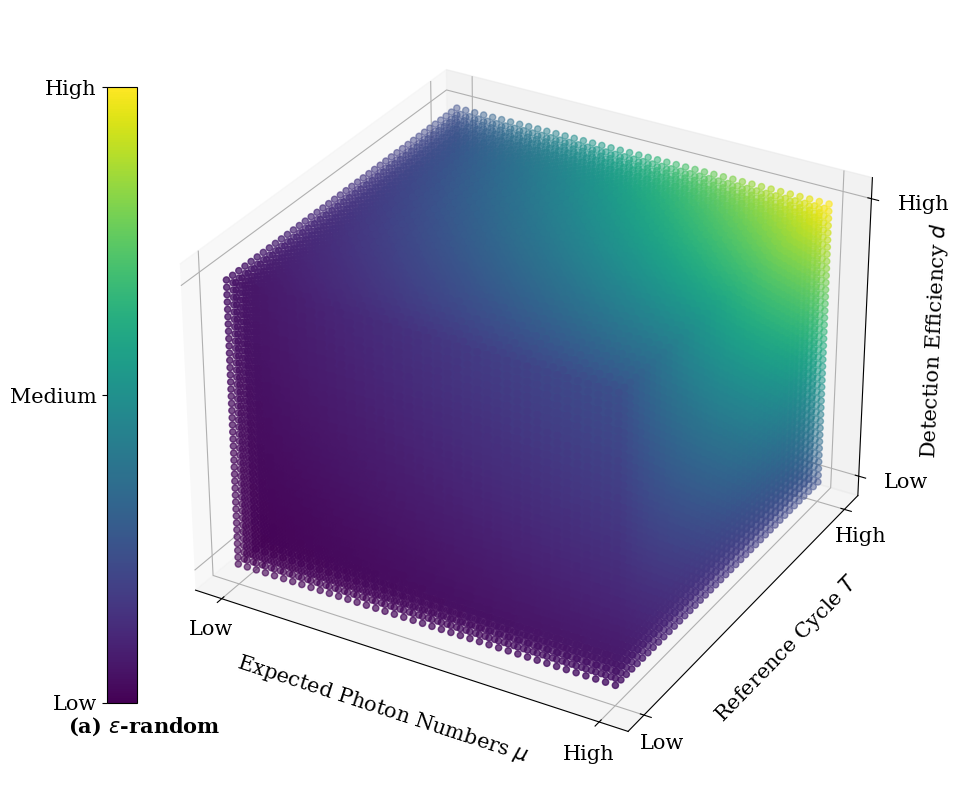}\hfill\includegraphics[width = 0.49\textwidth]{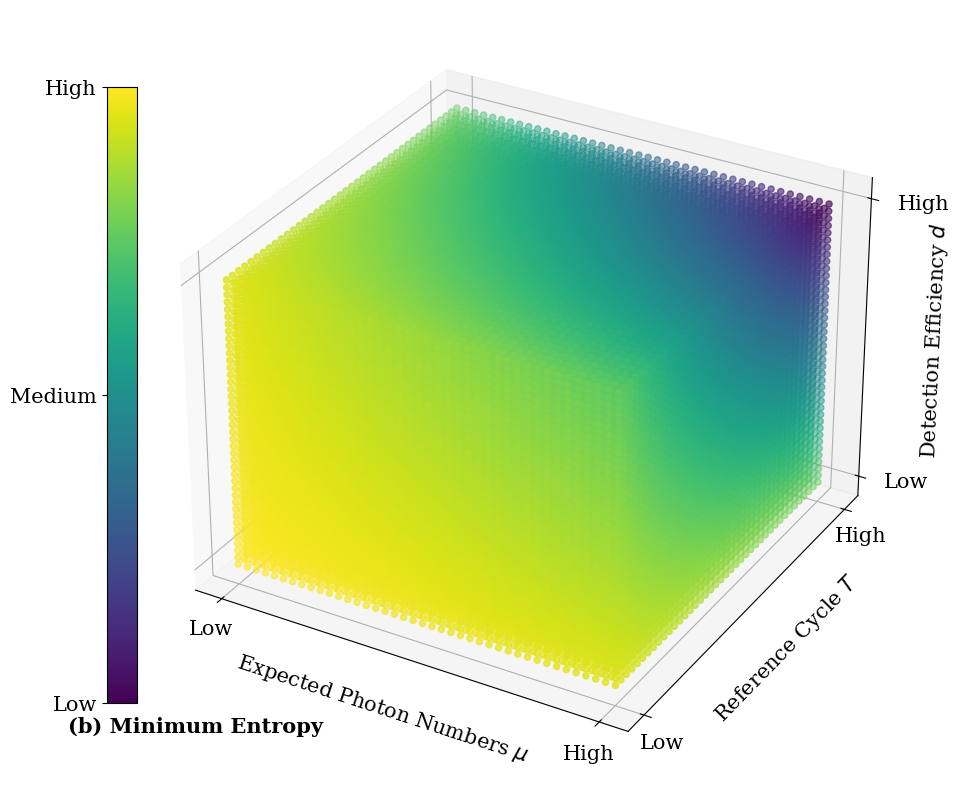}\\
\vspace{0.1in}
\includegraphics[width=0.49\textwidth]{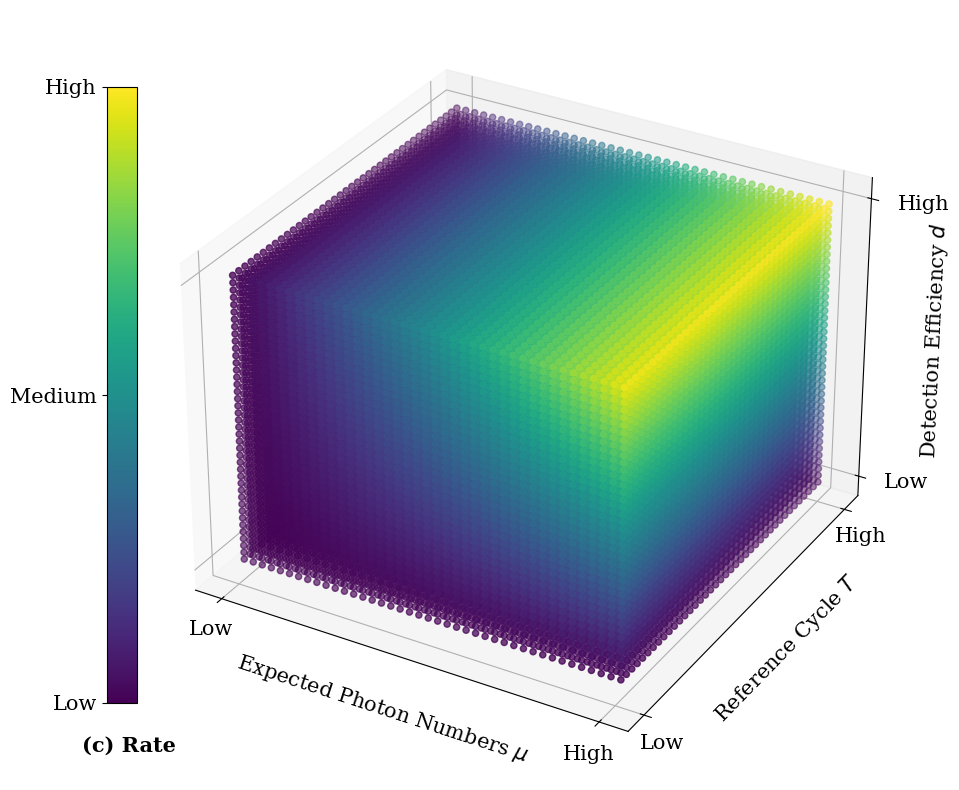}\hfill\includegraphics[width=0.49\textwidth]{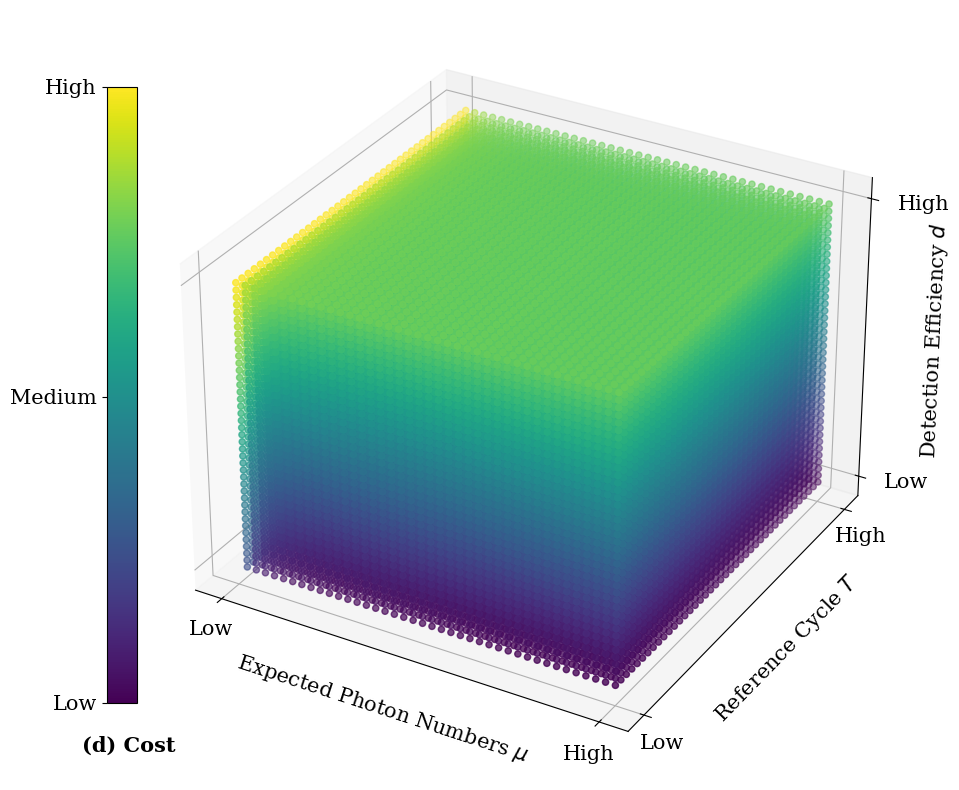}
\caption{Heat map denoting how (a) $\epsilon$-randomness, (b) minimum entropy, (c) random number generation rate and (d) cost varies depending on the expected photon number per unit time, $\mu$, external reference cycle, $T$ and the detection efficiency, $d$. Figure (a) and (b) show the $\epsilon$-randomness in Eq.~\eqref{E:randomness_muTd} and minimum entropy in Eq.~\eqref{E:min_ent}, respectively, indicating that we must keep the product $\mu Td$ small to get high randomness. However, (b) for Eq.~\eqref{E:murate}, with $N=256$, shows if the $\mu d$ is small, the generation rate will be minimal. On the other hand, from (c) for Eq.~\eqref{E:cost} with $\alpha=1/20, \beta=40$, low-cost demands a small value of $d$ and a high value of $\mu$.}
\label{fig:h_map}
\end{figure*}

The above discussion has been described as a heat map in Fig.~\ref{fig:h_map}. In the figure, dark shade denotes low value and light shade denotes high value of $\epsilon$, minimum entropy, generation rate and cost in Figures~\ref{fig:h_map}(a),~\ref{fig:h_map}(b),~\ref{fig:h_map}(c) and~\ref{fig:h_map}(d), respectively. Figure~\ref{fig:h_map}(a) and (b) shows that high randomness demands a small value of the product $\mu Td$. However, the product $\mu d$ cannot be very small as Fig.~\ref{fig:h_map}(b) indicates. A small $\mu d$ would result in a low random number generation rate. On the other hand, to keep the cost of the QRNG low, we need a small $d$ but a large $\mu$.

In Table~\ref{tab:mutd}, we summarize the discussion in tabular form. A down-arrow ($\downarrow$) indicates a decrease in the value of the corresponding quantity, while an up-arrow ($\uparrow$) signifies an increase. The external reference cycle $T$ does not influence speed or cost. If the parameters are increased, the effects will reverse accordingly.

We have already mentioned in Eq.~\eqref{E:goal_ent} that the main goal in designing a QRNG is to keep $H(Q)$, the entropy from the quantum source as close as $H(U)$, the entropy of uniform randomness. However, we also discussed that this is not the only goal. There are some other constraints as well like random number generation rate and cost. A QRNG should generate random numbers with high rates and low costs. The randomness and the rate together describe the \emph{performance} of the QRNG. Therefore, the goal in designing a QRNG is to maximize the performance and minimize the cost by choosing appropriate $\mu, T$ and $d$.

\section{\label{sec:NIST}Results of Randomness Tests}

\begin{figure*}
\centering
\includegraphics[width=0.95\columnwidth]{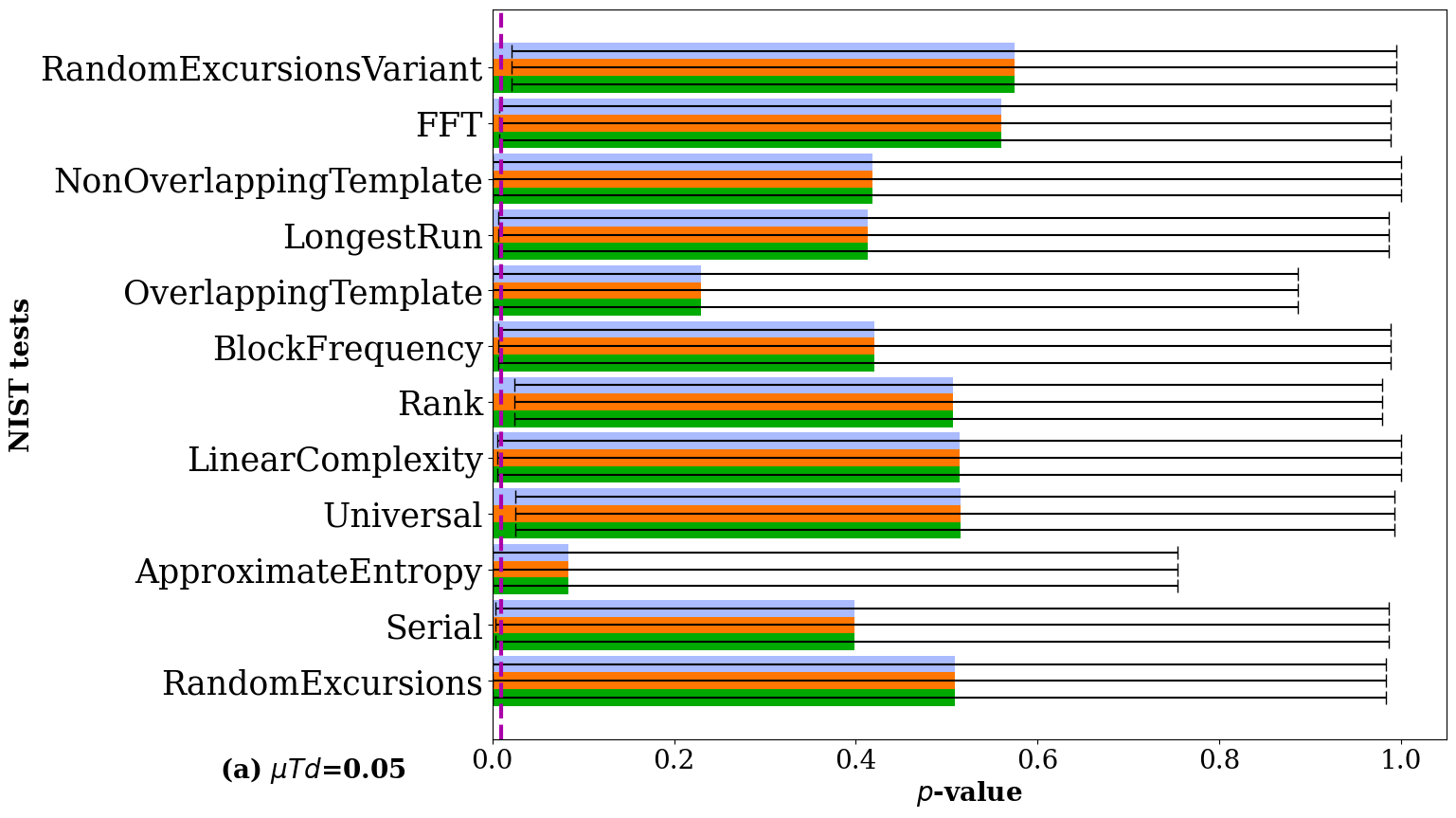}
\includegraphics[width=0.95\columnwidth]{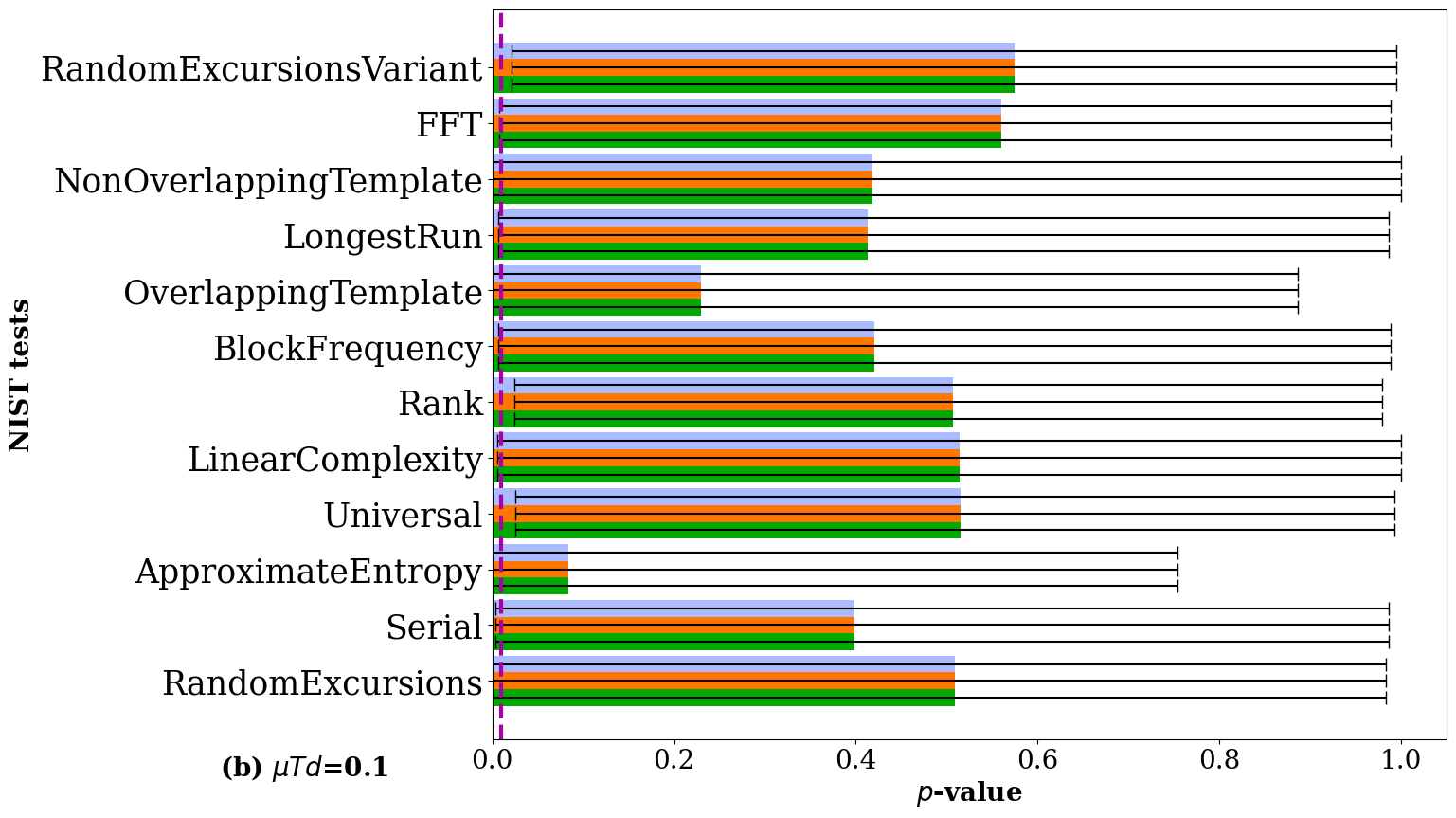}
\includegraphics[width=0.95\columnwidth]{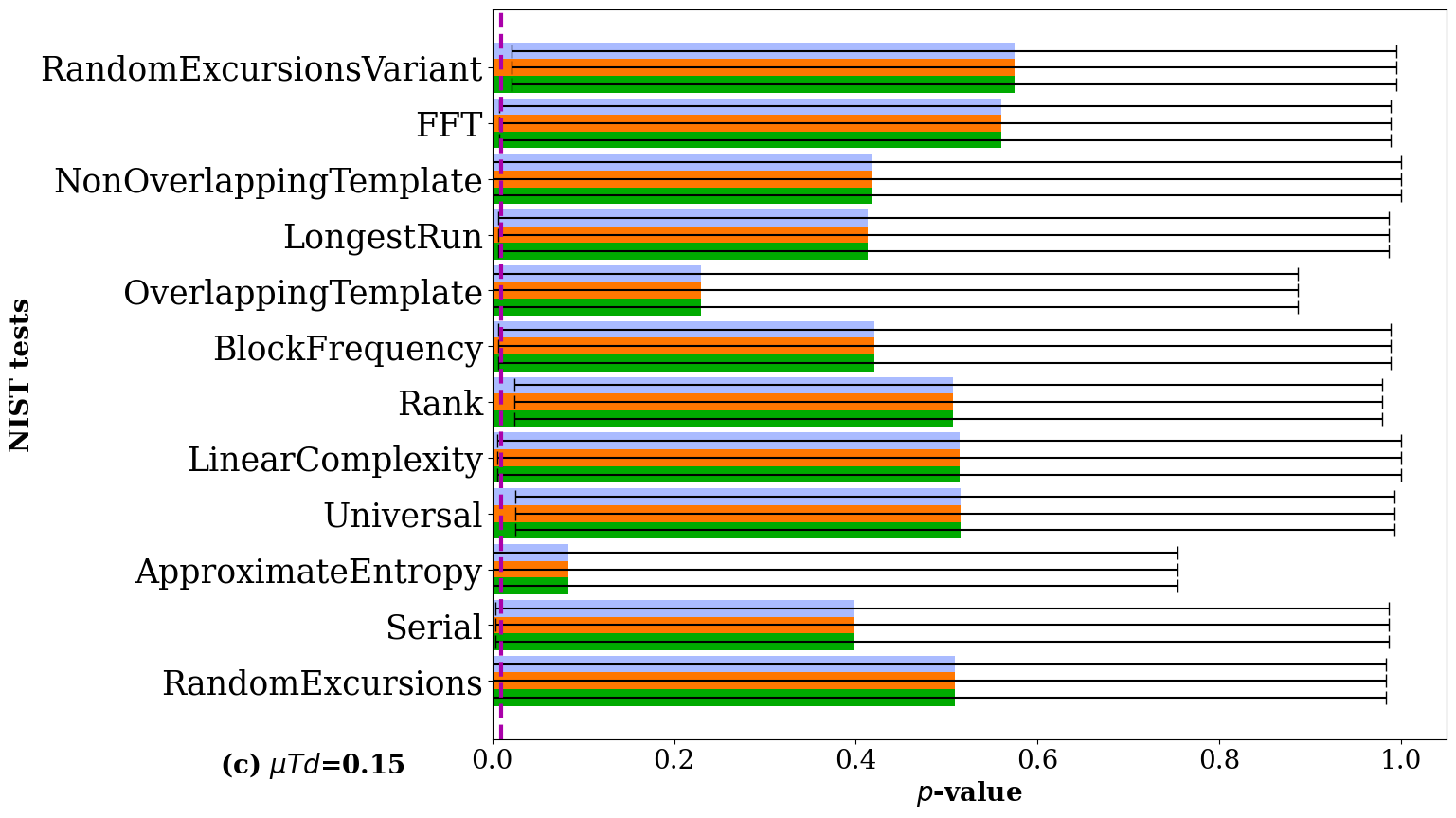}
\includegraphics[width=0.95\columnwidth]{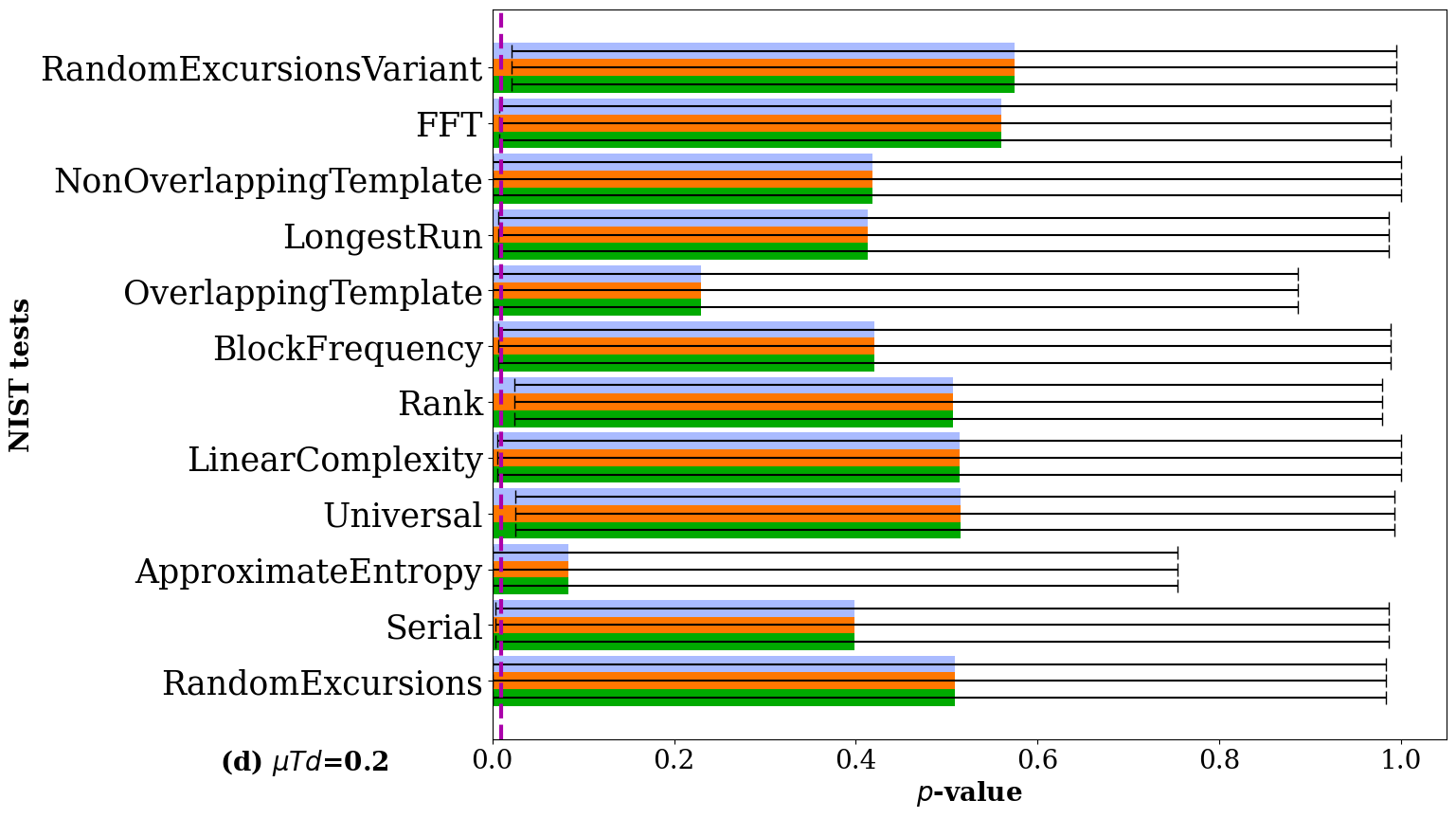}
\includegraphics[width=0.95\columnwidth]{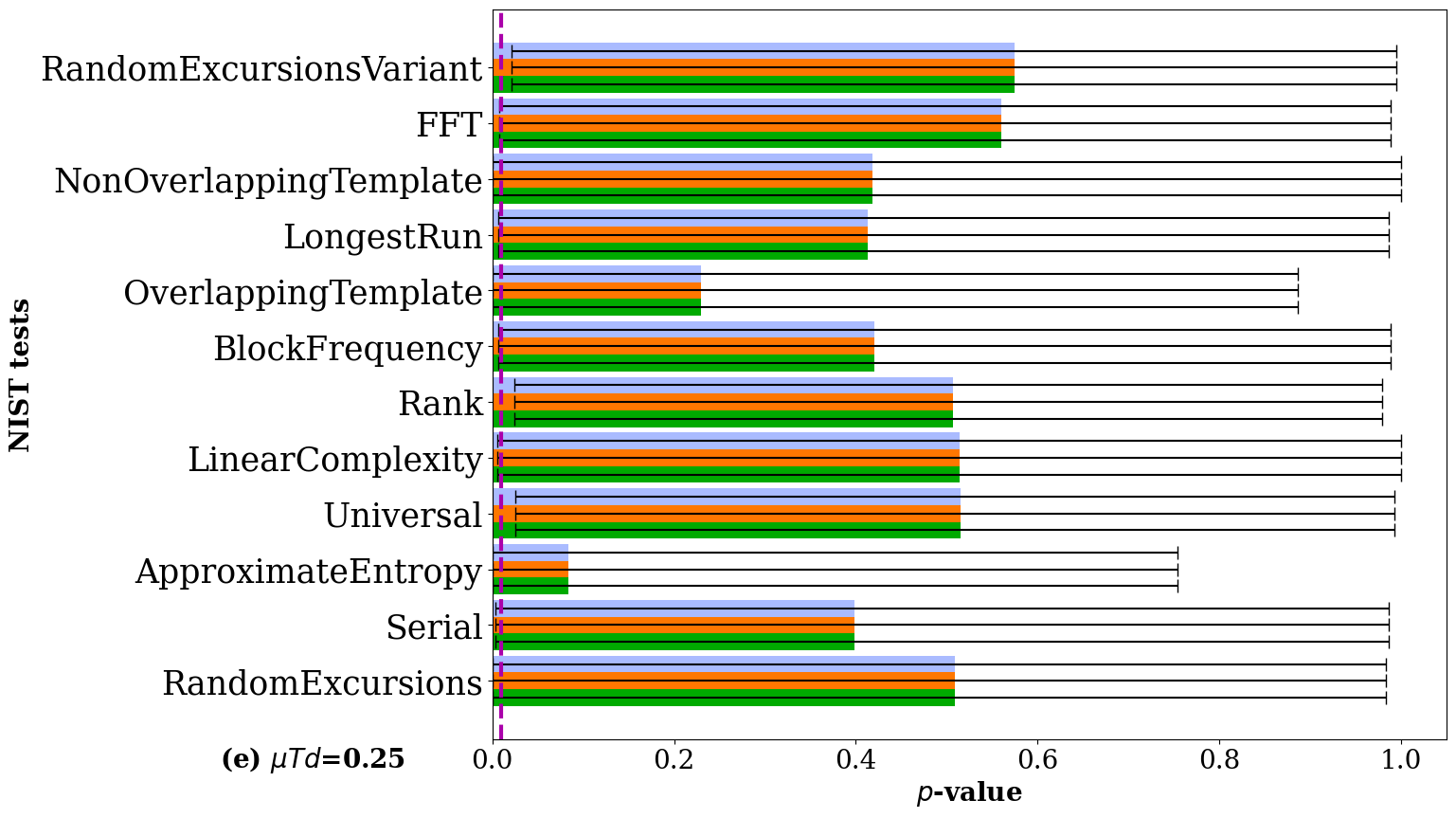}
\includegraphics[width=0.95\columnwidth]{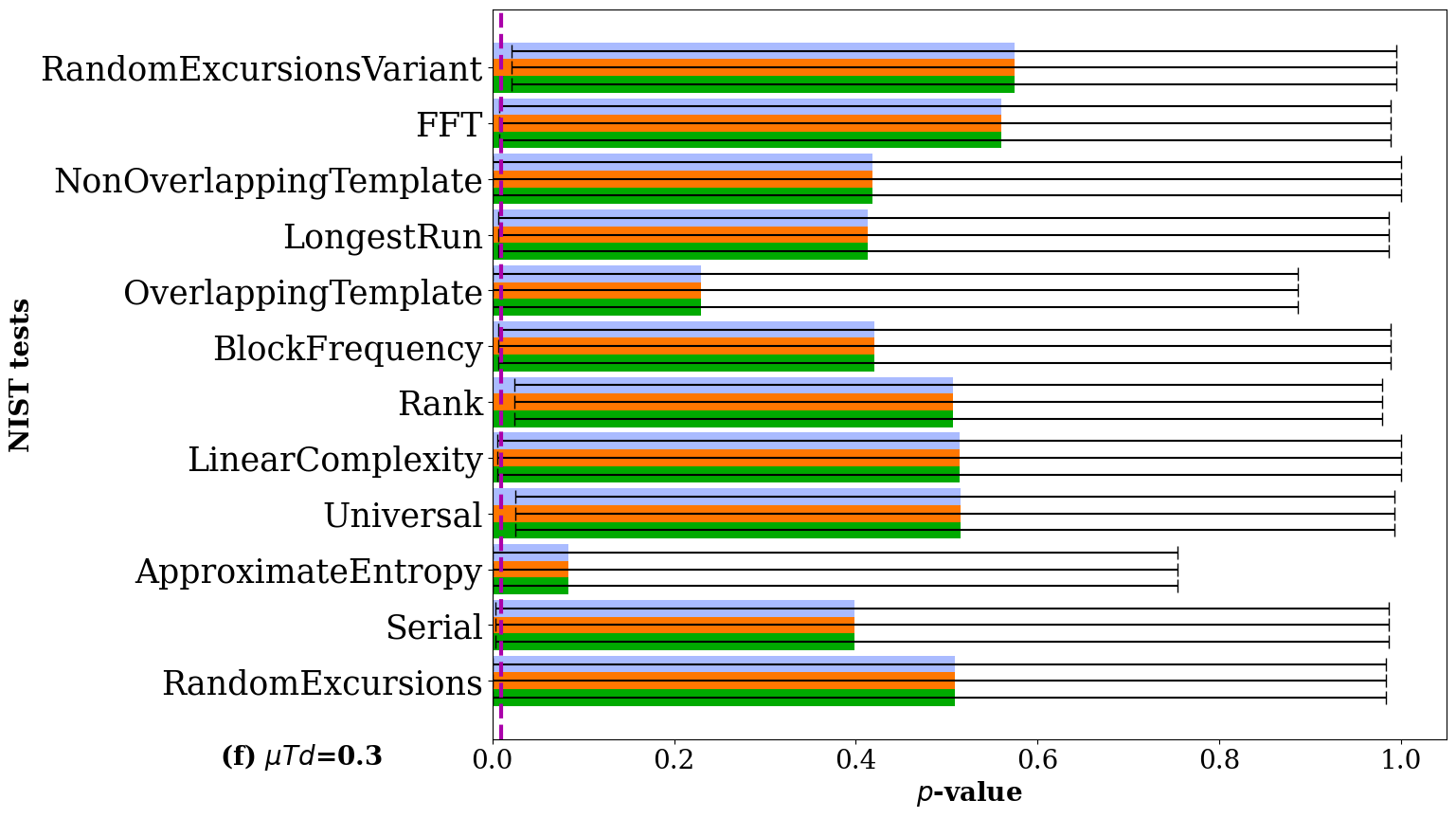}
\includegraphics[width=0.95\columnwidth]{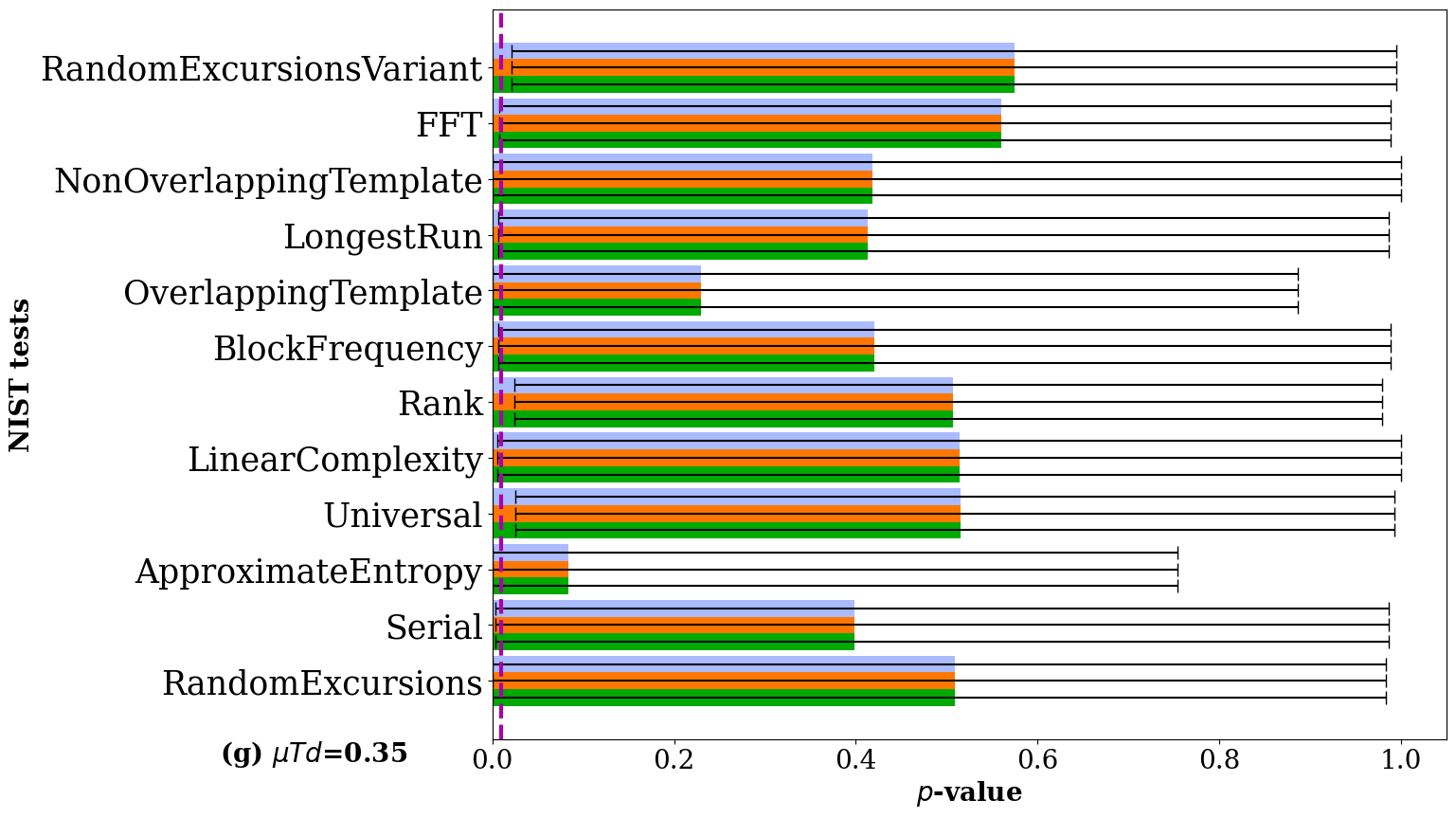}
\includegraphics[width=0.95\columnwidth]{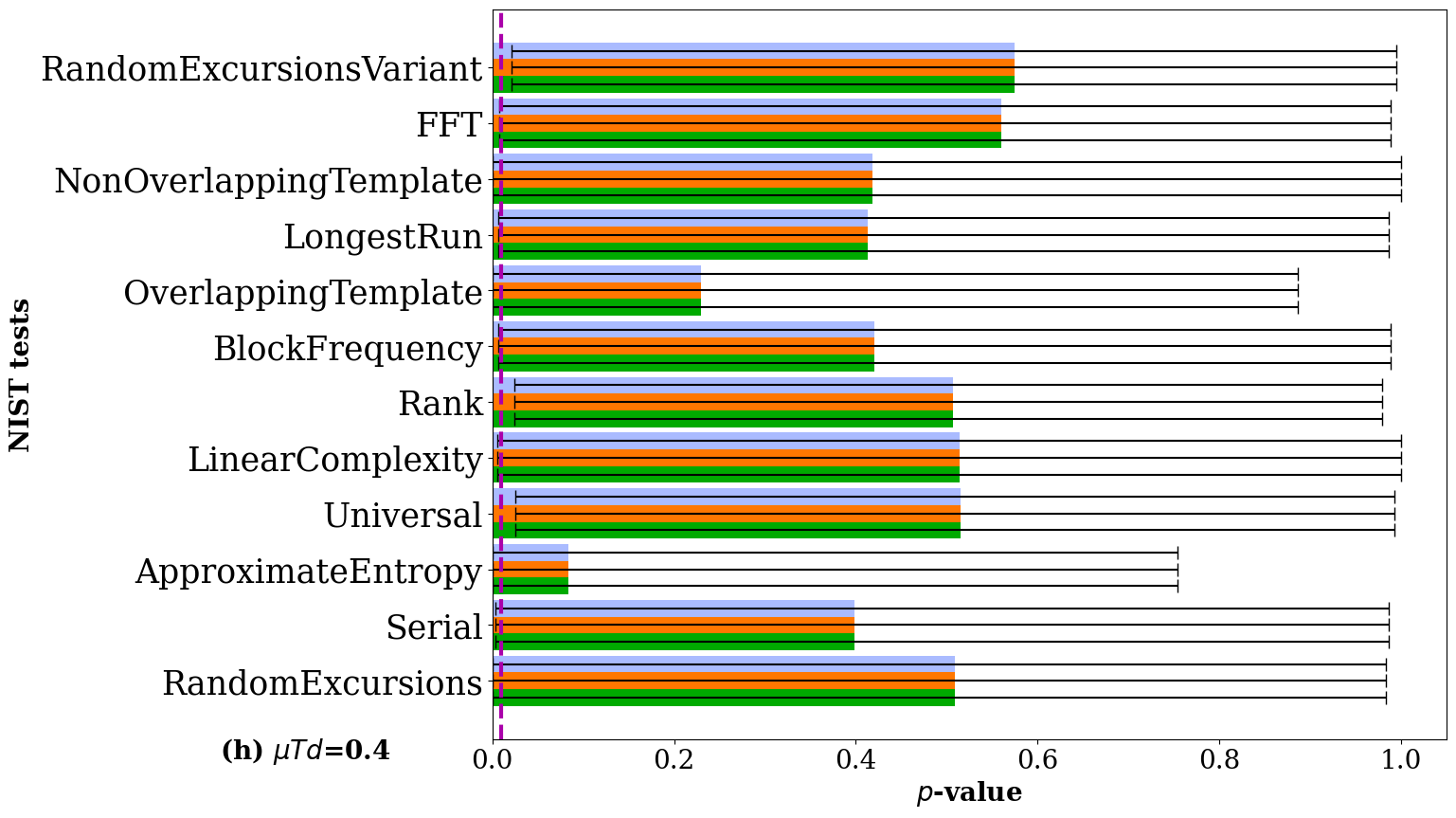}
\caption{\label{fig:nist_pval}NIST result. $p$-values for passed tests are plotted as bar charts for (a) $\mu Td=0.05$, (b) $\mu Td=0.10$, (c) $\mu Td=0.15$, (d) $\mu Td=0.20$, (e) $\mu Td=0.25$, (f) $\mu Td=0.30$, (g) $\mu Td=0.35$, (h) $\mu Td=0.40$.
For each $\mu Td$, three $1$ GB files are generated by simulating probability distribution~\eqref{E:distIII}, and the NIST tests have been performed on those files. Different textures denote results corresponding to different files. A test is considered as passed if the corresponding $p$-value is at least $0.01$. Tests that fail for all of the three files are not included in the bar chart. The vertical dashed line denotes $p$-value $=0.01$}
\end{figure*}

Random numbers and their randomness testing have been discussed in~\ref{sec:SPQRNG}. The randomness of the random numbers is tested using statistical tests. Here we used four common statistical testing suites namely NIST, Dieharder, AIS-31 and ENT.

\begin{figure}
\centering
\includegraphics[width=\columnwidth]{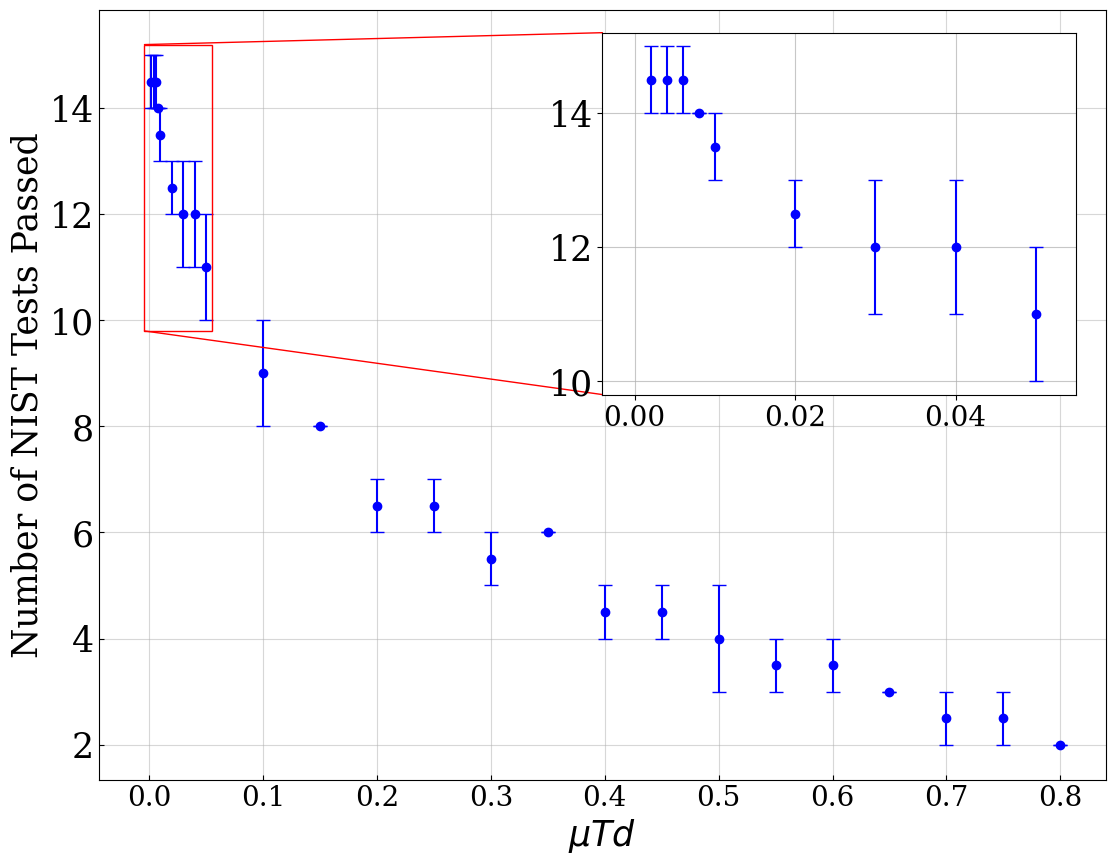}
\caption{NIST test results for the data simulated with the distribution given by Eq.~\eqref{E:distIII}. Three files, each containing $1$ GB random number have been used for the test for every $\mu Td$. A test fails if the corresponding $p$-value is found less than $0.01$. The range in the second column arises due to different results from three different files. The number of passed tests decreases with the increment of $\mu Td$. This indicates that as $\mu Td$ increases, the randomness in the generated random number decreases.}
\label{fig:nist}
\end{figure}

\emph{The NIST Randomness Testing Suite} is an extensive set of statistical tests developed by the National Institute of Standards and Technology (NIST) to evaluate the quality of randomness in random numbers. It is one of the most widely used suites for testing randomness. It contains $15$ different tests to examine different aspects of randomness. Most of the tests have the standard normal or $\chi^2$ as reference distribution for their test statistic. Here null hypothesis is taken as \emph{the number to be tested is random.} For each test, the significance level is considered as $0.01$. If the value of the test statistic for some tests falls in the critical region, or equivalently, calculated $p$-value is less than $0.01$, then the corresponding tests fail rejecting the null hypothesis. A perfectly random number should pass all of the $15$ tests. As mentioned in Section~\ref{sec:2fold}, the central limit theorem is applicable only for large $n$, tests that use the central limit theorem to determine the reference distribution, have a requirement for the bit stream, that is, the length of the testing sequence, to be large enough (at least $10^6$).

\begin{figure}
\centering
\includegraphics[width=\columnwidth]{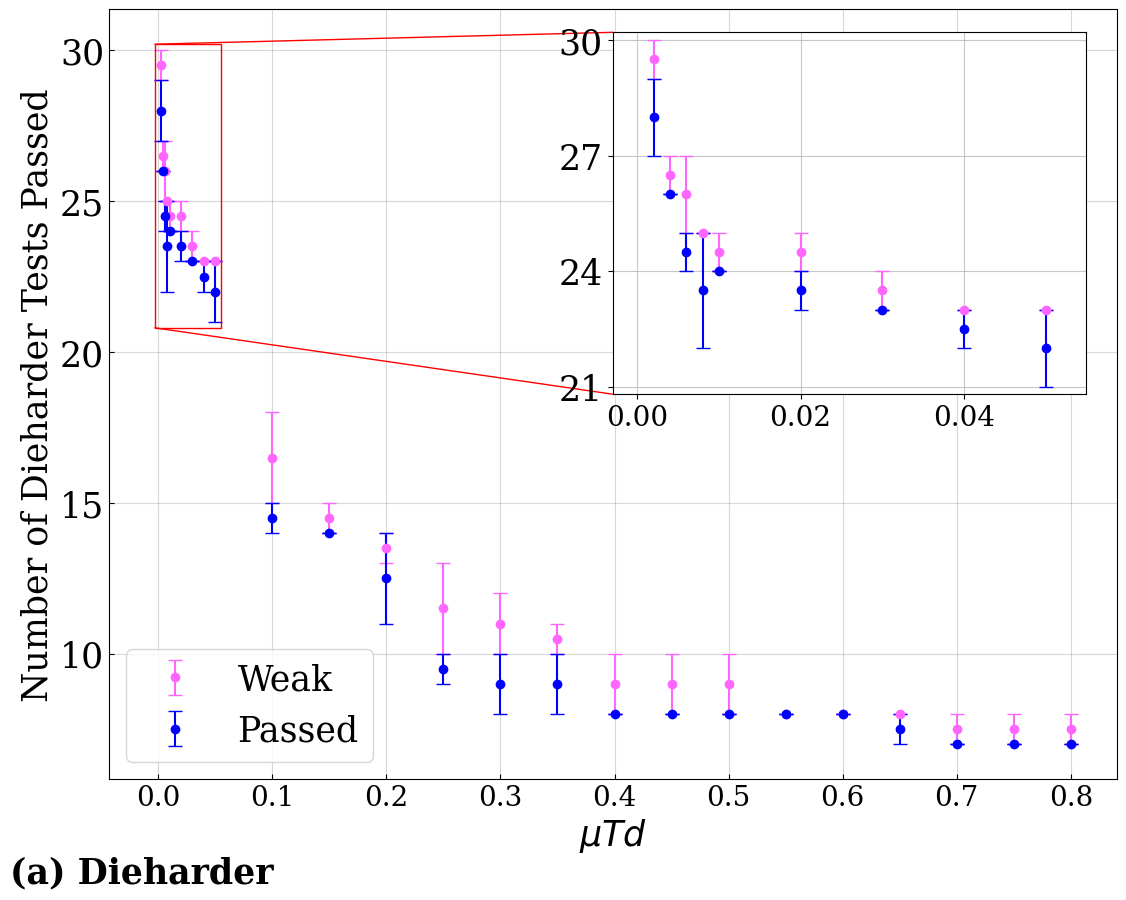}\hfill
\includegraphics[width=\columnwidth]{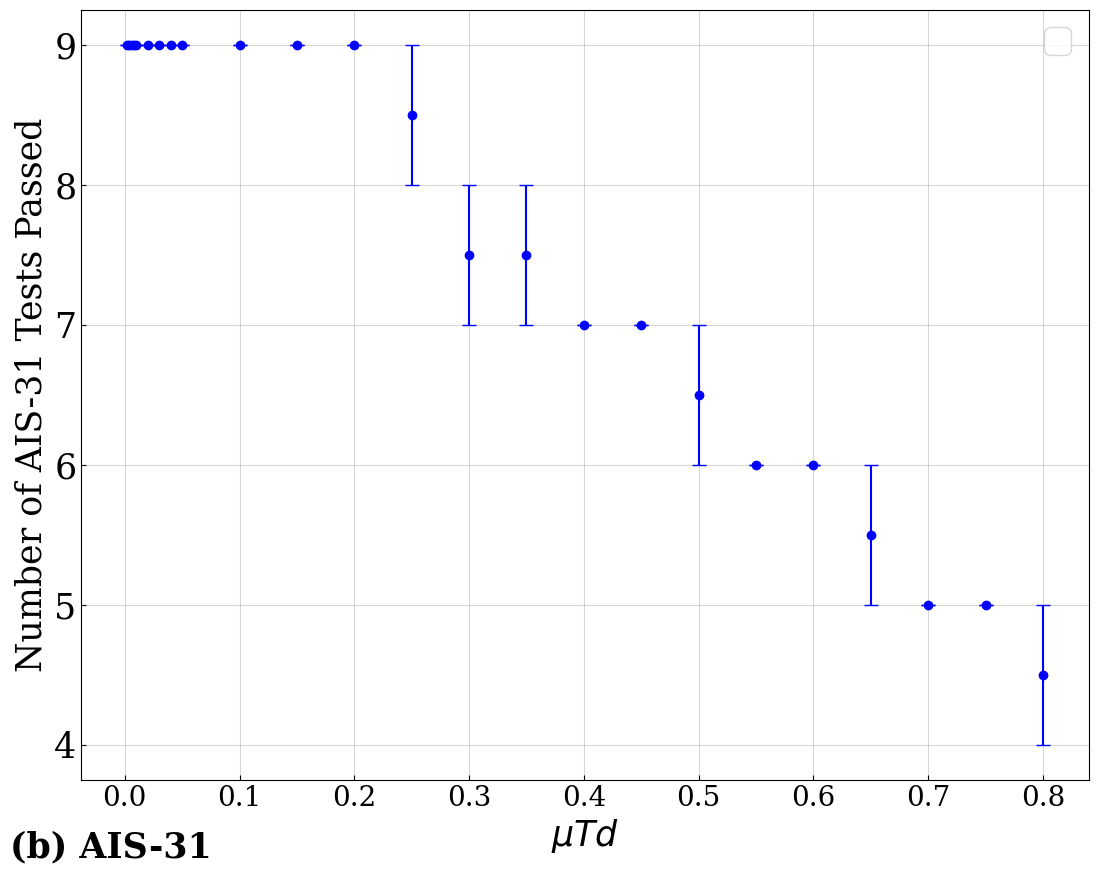}
\caption{(a) Dieharder and (b) AIS-$31$ test results for the data simulated with the distribution given by Eq.~\eqref{E:distIII}. For Dieharder, there are $30$ tests; for AIS-$31$, $9$ tests are there. These tests also reveal that the number of passed tests decreases as $\mu Td$ increases, indicating that lower values of $\mu Td$ result in better randomness before post-processing. Along with pass and fail, Dieharder tests also mention suspected results as weak. They are also included in this figure.}
\label{fig:Die-AIS}
\end{figure}

\emph{The Dieharder} tests represent a battery of statistical tests used to assess the quality and randomness of a RNG. This suite contains around $30$ tests for the assessment. These tests consider KS test of uniformity to obtain the $p$-values. Depending on these $p$-values, the decision (\textit{passed}, \textit{weak} or \textit{failed}) has been taken. The results that do not pass and also do not have enough evidence to declare as fail have been mentioned as weak.

\emph{AIS-$31$} is a standardized suite of statistical tests for assessing RNGs. This suite comprises a total of 9 tests, typically organized into two main groups. Procedure A, consists of 6 tests aimed at detecting statistically inconspicuous behavior in RNG output. Procedure B, focuses on evaluating the internal random bits of TRNGs through 3 specific tests. These tests ensure that RNGs meet rigorous statistical criteria for randomness, providing a reliable measure of their quality and suitability for applications requiring secure and unpredictable random numbers.

\begin{figure}
\centering
\includegraphics[width=\columnwidth]{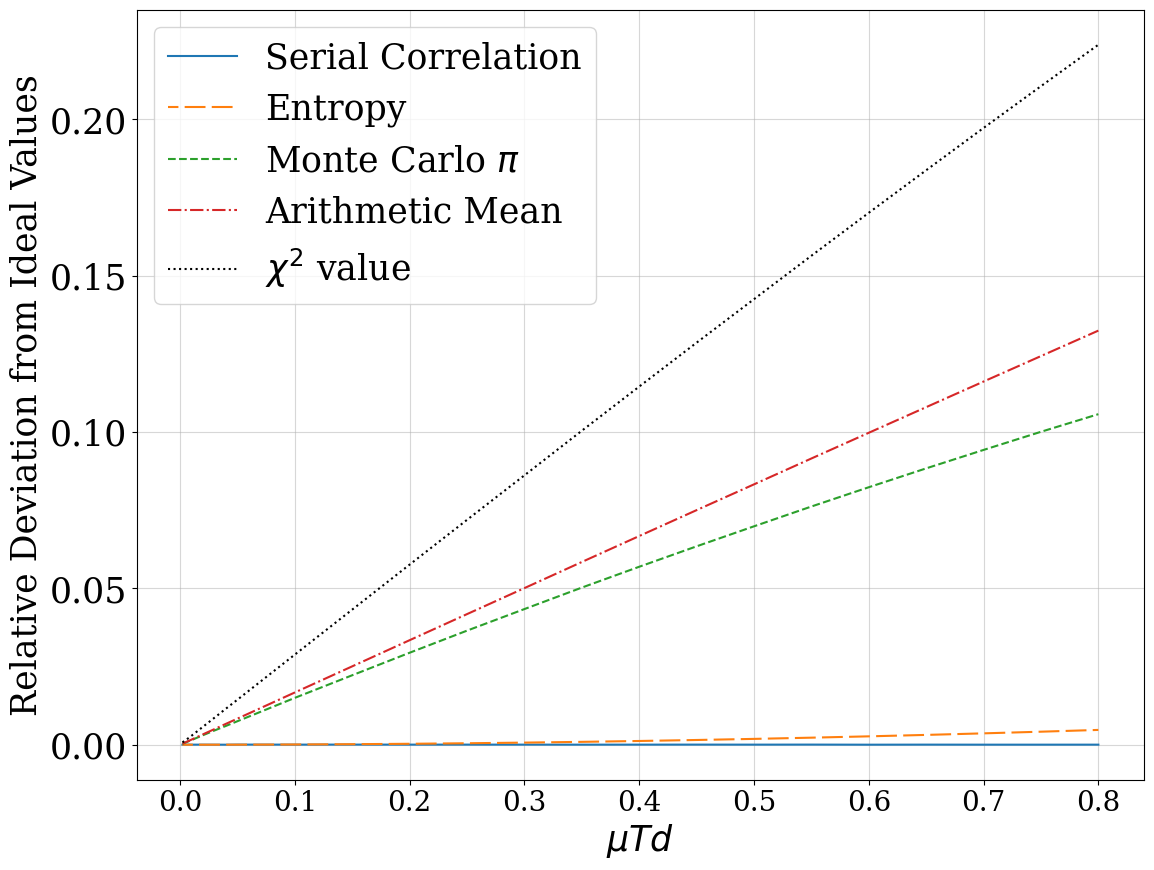}
\caption{Ent test result for the data simulated with the distribution given by Eq.~\eqref{E:distIII}. The relative deviation is calculated as $\frac{|v_i-v_o|}{v_i}$, where $v_i$ is the expected ideal value for uniform distribution and $v_o$ is the observed value for the simulated data. Although the deviation of the Monte Carlo value of $\pi$, the arithmetic mean and the $\chi^2$ test value is high, the deviation of the entropy and the serial correlation is very small.}
\label{fig:ENT_res}
\end{figure}

\emph{ENT} performs $5$ tests on the stream of bytes and produces the outputs for each test. It calculates entropy, $\chi^2$ statistic value, the arithmetic mean of the bytes, the Monte Carlo value of $\pi$ and the serial correlation coefficient. Depending on these values user can decide the randomness.

We have generated random numbers simulating the probability distribution given by Eq.~\eqref{E:distIII} for different $\mu Td$. Three files of size $1$ GB have been generated for every $\mu Td$. Then we perform the NIST tests over the generated numbers. $100$ bit streams of length $10,000,000$ have been considered for the tests. In Fig.~\ref{fig:nist_pval}, we plot the bar charts of the $p$-values from different NIST tests. The tests for which the $p$-values are found as $0$ are not included in the plot. From these plots, we can see that, as $\mu Td$ increases, the number of passed tests decreases, indicating a decrement of the randomness in the random number. The overall result is plotted in Fig.~\ref{fig:nist} as the number of tests our simulated data passed. These graphs indicate that lower $\mu Td$ values give rise to better random numbers requiring less post-processing.

We have also performed the Dieharder, AIS-$31$ and ENT over the generated numbers. Figure~\ref{fig:Die-AIS} shows how many tests for our simulated data passed for Dieharder and AIS tests. Similar to the NIST tests, these tests also show that small $\mu Td$ values provide better randomness before post-processing. ENT test does not mention pass or fail as a result. However, it provides ideal and actual values for the performed tests. In Fig.~\ref{fig:ENT_res}, we plot the relative deviation of the actual values from their ideal values for different ENT tests. In this case also, the deviations increase with the increase of $\mu Td$.

\section{\label{sec:conc}Discussion}

In this paper, we consider two perspectives for analyzing QRNG. From the manufacturer's point of view, we discuss how to determine when the source can be considered as quantum. Here we assume that photon detection time is available as a sample. We suggest two-fold statistical tests. The first one may be an interval estimation or hypothesis test to decide between sub-Poissonian and super-Poissonian. On the other hand, the last one, which is a $\chi^2$-test ensures Poisson distribution. This test can be performed for any quantum-photonic device to check its quantumness.

Then we discuss photon statistics for time-bin-based QRNG in practical scenarios. We consider two photon-arrival-time-based QRNG models, one using external time reference and another without such external reference. We show that, although for ideal devices these two models follow different distributions (one of which is perfectly random), these two QRNGs follow a similar distribution when the photon source is not an actual single photon source. We derive a relation when the photon statistics becomes $\epsilon$-random. We found that by making the product of the expected number of photons ($\mu T$) within an external reference cycle ($T$) and the detection efficiency ($d$) small enough, the randomness of the generated raw random number can be increased. Therefore it would require less post-processing, and it would retain maximum randomness from the quantum source. This discussion would help QRNG manufacturers to design better QRNG.

Although small $\mu Td$ provides better randomness, from the user's perspective it may not be suitable. The random number generation rate, the timing error in registering photon detection time and the cost of the QRNG restrict us from choosing small values for $\mu, T$ or $d$. A lower expected photon count improves quantumness but comes with increased costs and a reduced generation rate. On the other hand, a shorter external reference cycle enhances quantumness but must surpass a minimum threshold to reduce timing errors, having only slight effects on cost and rate. Reduced detection efficiency boosts quantumness and lowers costs, but also decreases the generation rate. The trade-off among the randomness sourced from quantum systems, the rate of random number generation, and the cost of the QRNG is contingent upon the specific requirements of the users.

Finally, we simulated random numbers according to the photon statistics we derived and applied statistical tests including NIST, Dieharder, AIS-31, and ENT to these numbers. The results from these tests validate our theoretical conclusions based on $\mu Td$ about the randomness of the quantum sources.

In practical cases, the dark count of a detector also contributes to the generation of random numbers. In our discussion, we did not consider the dark count. This work can be extended to see the effect of dark counts on the photon statistics. Here, we have considered two models of photon arrival-time-based QRNGs, which fall in the category of trusted devices. A similar analysis can be performed for other QRNG models.

One may note that our discussions are not meant for black-box testing for commercially available QRNGs. Rather, the physicists and engineers may use these in the QRNG design phase before post-processing, for evaluation, fine-tuning and performance optimization.

\section*{Acknowledgement}
S. Das would like to acknowledge the support received by the Quantum Delta NL KAT-2 project.

\section*{Data Availability}

Datasets that have been used to validate the two-fold method and the Python codes which are used to generate these datasets are available at\\ \href{https://github.com/sp-k/QRNG_eval.git}{https://github.com/sp-k/QRNG\_eval.git}.

\appendix

\section{\label{sec:g2t}Relationship between Non-classicality of Light, antibunching of Photons and sub-Poissonian Distributions}

The distribution of the photon-counting statistics can be divided into three classes~\cite{fox2006quantum}:
\begin{enumerate}
\item[(i)] sub-Poissonian, i.e., the distribution variance is less than the mean,
\item[(ii)] Poissonian, i.e., the distribution variance is the same as the mean, and
\item[(iii)] super-Poissonian, i.e., the distribution variance is greater than the mean.
\end{enumerate}
Classical wave theory of light follows Poissonian and super-Poissonian statistics~\cite{fox2006quantum}. On the other hand, the sub-Poissonian nature of a photon source cannot be explained by the classical wave theory, making it a clear indication of its quantum nature~\cite{RevModPhys.68.127,qute.202100062}.

For example, thermal light follows Bose–Einstein distribution~\cite{fox2006quantum}, given by
\begin{equation}
P(n)=\frac{1}{\mu+1}\left(\frac{\mu}{\mu+1}\right)^n.
\end{equation}
The mean of this distribution is given by $\mu$, and the variance is $\mu+\mu^2>\mu$. Therefore, this is a super-Poissonian distribution and hence thermal light is an example of classical light.

Next consider a perfect single-photon source that emits photons maintaining equal time interval, $\Delta t$ between two consecutive photons. Then the photon count within time $T$ would be the integer $\mu=\lfloor\frac{T}{\Delta t}\rfloor$. In this case, the mean photon count is given by $\mu$, but the variance is $0<\mu$. Therefore, this is an example of non-classical or quantum light.

A coherent photon source produces coherent states which can be represented as a superposition of Fock states by
\begin{equation}
\label{E:coh_st}
|\psi_\alpha\rangle=e^{-\frac{|\alpha|^2}{2}}\sum_{n=0}^\infty\frac{\alpha^n\left(\hat{a}^\dagger\right)^n}{\sqrt{n!}}|0\rangle,
\end{equation}
where $\alpha$ is a complex number and $\hat{a}^\dagger$ is the creation operator. The mean photon number corresponding to this state is given by $|\alpha|^2$. This state can be approximated as a single-photon state with high attenuation.  The probability of $n$ photons within $T$ time follows Poisson distribution~\cite{PhysRev.145.1041} given by
\begin{equation}
\label{E:Psn}
P(n,T)=\frac{e^{-\mu T}(\mu T)^n}{n!},
\end{equation}
where $\mu$ is the mean number of photons per unit time. It can be shown that for this distribution, the variance of the number of photons per unit time is also $\mu$.

Antibunching property can also be used to distinguish quantum light from classical light. The second-order correlation function~\cite{fox2006quantum} of light is defined by
\begin{equation}
\label{E:g2t}
g^2(\tau):=\frac{\langle I(t)I(t+\tau)\rangle}{\langle I(t)\rangle\langle I(t+\tau)\rangle},
\end{equation}
where $I(t)$ is the intensity of the light at time $t$ and $\langle\cdot\rangle$ denotes the average over all $t$. Using the classical theory of light, it can be shown that~\cite{RevModPhys.68.127}
\begin{equation}
\label{E:gc}
\begin{aligned}
g^2(0)&=\frac{\langle I(t)^2\rangle}{\langle I(t)\rangle^2}\geq1,\\
\text{and }g^2(\tau)&=1\text{ for }\tau\gg\tau_c,
\end{aligned}
\end{equation}
$\tau_c$ being the coherence time of the source.

Now, consider that a photon stream is sent through a beam splitter and two detectors $D_0$ and $D_1$ are kept in two output paths. Than the correlation function~\eqref{E:g2t} can be written as
\begin{equation}
\label{E:g2tc}
g^2(\tau):=\frac{\langle n_1(t)n_2(t+\tau)\rangle}{\langle n_1(t)\rangle\langle n_2(t+\tau)\rangle},
\end{equation}
where $n_i(t)$ is the number of counts registered on detector $i$ at time $t$. If the incoming light consists of a photon stream with long intervals between two consecutive photons, the product $n_1(t)n_2(t)$ is always $0$, giving $g^2(0)=0<1$, which is not possible with classical lights. Thus, depending on $g^2(t)$, light is distinguished in the following three classes~\cite{fox2006quantum}:
\begin{enumerate}
\item[(i)] bunched light, when $g^2(0)>1$,
\item[(ii)] coherent light, when $g^2(0)=1$,
\item[(iii)] antibunched light, when $g^2(0)<1$.
\end{enumerate}
Out of these three, antibunched light is undoubtedly non-classical.

The relation between photon count distribution and $g^2(\tau)$ is given by~\cite{PhysRevA.41.475}
\begin{equation}
\label{E:mvg}
\mean{(\Delta N)^2}-\mean{N}=\mean{N}^2\frac{1}{T^2}\int_{-T}^Td\tau(T-|\tau|)(g^2(\tau)-1).
\end{equation}
In the above equation, if $g^2(\tau)\leq1$ for all $\tau$, it implies $\mean{(\Delta n)^2}<\mean{N}$, i.e., a sub-Poissonian statistics, indicating non-classicality of light. However, if $g^2(\tau)\leq1$ for some $\tau$, there is no direct relation between the sub-Poissonian photon count statistic and the antibunching of photons. Xou and Mandel~\cite{PhysRevA.41.475} showed that if a source emits photons in two different nonvacuum modes, both having occupation number $n/2$, but different frequencies $\omega_1$ and $\omega_2$, the second-order correlation function takes the form
\begin{equation}
\label{E:g2t_counter}
g^2(\tau)=\frac{1}{2}\cos(\omega_1-\omega_2)\tau-\frac{1}{n}+1,
\end{equation}
for $\tau<T$. Also, if $P(N,t,t+T)$ be the distribution of photon count in the time interval from $t$ to $t+T$,
\begin{equation}
\label{E:Photo_counter}
\mean{(\Delta N)^2}-\mean{N}=\mean{N}^2\left[\left(\frac{\sin(\omega_1-\omega_2)T/2}{(\omega_1-\omega_2)T/2}\right)^2-\frac{1}{n}\right],
\end{equation}
where $\mean{N}$ and $\mean{(\Delta N)^2}$ are respectively the mean and the variance of the above distribution. Choosing $T=\frac{2\pi}{\omega_1-\omega_2}$, it can be seen that $\mean{(\Delta N)^2}<\mean{N}$, i.e., the light is sub-Poissonian although $g^2(0)=(\frac{3}{2}-\frac{1}{n})>1$ for $n>2$ denoting the light is bunched. This example also shows that although both the antibunching of photons and the sub-Poissonian distribution of photon count indicate the non-classicality of light, they are not identical.

\section{\label{sec:DIQRNG}A Comparison between trusted QRNG and DI-QRNG}

\begin{table*}[htpb]
\centering
\begin{tabular}{|c|p{0.4\textwidth}|p{0.4\textwidth}|}
\hline
&\textbf{Trusted QRNG}&\textbf{Device-independent QRNG}\\
\hline
\multirow{5}{0.175\textwidth}{\textbf{Advantages of DI-QRNG}}&1. Required trusted assumptions.&1. Trusted assumptions are not required or partially required for semi-DI-QRNG.\\
\cline{2-3}
&2. Security relies on trusted assumptions. Hence, less secure.&2. Security relies on Bell violation or similar quantum principles~\cite{acin2016certified,Alicki_2008,Alicki_2008a,Martin_2023}. Hence, more secure.\\
\hline
\multirow{14}{0.125\textwidth}{\textbf{Advantages of Trusted QRNG}}&1. No entanglement required. Hence it is cheaper.&1. Necessity of entanglement makes it costly.\\
\cline{2-3}
&2. No Bell test required.&2. Loophole-free Bell test is needed and performing a loophole-free Bell test is difficult~\cite{PhysRevA.47.R747,GISIN1999103,PhysRevA.66.042111,PhysRevA.81.032106,Larsson_2014}.\\
\cline{2-3}
&3. Can be produced in the very compact form of USB or chip.&3. To avoid locality loophole in the Bell test, two entangled particles must be space-like separated~\cite{PhysRevLett.115.250402,storz2023loophole}. This forces the device to be either large or trusted during the Bell test.\\
\cline{2-3}
&4. Compatible for portable devices.&4. Incompatible in use cases with portable devices, e.g., automobiles, satellites, point of sale (POS) terminals, portable phones, laptops, etc.~\cite{quantis}\\
\cline{2-3}
\hline
\end{tabular}
\caption{A comparison between trusted and Device-independent QRNGs indicating their advantages and disadvantages.}
\label{tab:comp_QRNG}
\end{table*}

A device-independent quantum protocol uses quantum entanglement to ensure security. This can be verified using Bell inequality, quantum discord, etc.~\cite{acin2016certified,Alicki_2008,Alicki_2008a,Martin_2023}. On the other hand, if a protocol relies on single photons as a resource, the absence of entanglement prevents the application of the validation procedures mentioned above. However, we can still verify its quantumness using photon-count statistics~\cite{PhysRevA.41.475,fox2006quantum,RevModPhys.68.127,qute.202100062}, discussed in~\ref{sec:g2t}, to improve the security. A detailed comparison indicating advantages and disadvantages is given in Table~\ref{tab:comp_QRNG}.

From Table~\ref{tab:comp_QRNG}, one can easily conclude that although DI-QRNG provides more security, its size is not scalable to integrate it in a small space. Thus it is not suitable for many application scenarios requiring portable devices. Also, the high cost of DI-QRNG compared to trusted QRNG makes it incompatible with some uses like lottery games, token generations, etc.

\section{\label{sec:stat_tools}Statistical Tools}

\subsection{\label{sec:pnt_est}Point Ectimation}

In statistics, a \emph{statistic} $a$, which is a function of sample values, is called an \emph{estimator} for a distribution \emph{parameter} $\Theta$, if $P(A=\Theta)=1-\epsilon$ for some given $\epsilon\in(0, 1)$, where $A$ is the random variable corresponding to the statistic $a$. The procedure of finding such an estimator is called \emph{point estimation}. The quantity $1-\epsilon$ is called \emph{confidence level} of the estimator. A statistic $a$ is called an \emph{unbiased estimate} of an unknown parameter $\Theta$ if the expectation $\langle A\rangle=\Theta$. A statistic $a$ is called a \emph{consistent estimate} of a parameter $\Theta$ if for every real $\epsilon>0$, $\lim\limits_{n\to\infty}P(|A-\Theta|\geq\epsilon)=0$, where $n$ is the sample size.

\subsection{\label{sec:int_est}Interval Estimation}

In statistics, an interval $(a,b)$ is called the \emph{confidence interval} for a distribution parameter $\Theta$, if $P(A<\Theta<B)=1-\epsilon$ for some predefined $\epsilon\in(0, 1)$, where $A,B$ are the random variables corresponding to the statistics $a,b$ (they are functions of sample values) respectively. The quantity $1-\epsilon$ is called \emph{confidence level} of the interval. The procedure of finding the confidence interval for the parameter $\Theta$ is called \emph{interval estimation}.

\subsection{\label{sec:hyp_test}Hypothesis Testing}

In statistics, a \emph{hypothesis} is a statement or an assumption about a population distribution, this may be right or wrong. \emph{Hypothesis testing} is a procedure to decide whether a statistical hypothesis, called the \emph{null hypothesis} $H_0$, should be rejected or not. A \emph{critical region} or \emph{rejection region} of hypothesis testing is a specific region of test statistic values, for which the null hypothesis is rejected. The \emph{test statistic} for a hypothesis test is a function of the testing sample~\cite{casella2002statistical}. A hypothesis test is called \emph{one-tailed} if the critical region appears only at one tail of the reference distribution of the test statistic. On the other hand, it is called a \emph{two-tailed} test, if the critical regions appear at both tails of the above distribution.

During hypothesis testing, two types of error may occur. \emph{Type I error} occurs when $H_0$ is rejected given that it is true. \emph{Type II error} occurs when $H_0$ is accepted given that it is false. Let $\alpha$ and $\beta$ be the probabilities of Type I and Type II errors, respectively. The value $\alpha$ denotes the area of the critical region and is also called the \emph{significance level}. The value $1-\beta$, that is, the probability that $H_0$ is rejected given that it is false, is also called the \emph{power of a test}.

A hypothesis test starts with two hypotheses: the null hypothesis ($H_0$) and the alternative hypothesis ($H_a$). Depending on the significance level $\alpha$, the critical region is decided. The acceptance or rejection decision of $H_0$ is based on what is called the \emph{$p$-value} of the test statistic. It is defined as $P(|X|>|x|)$, where $X$ is a random variable, whose distribution is the reference distribution of the test statistic of a hypothesis test, and $x$ is the observed value of this statistic~\cite{ross2009introduction,hogg2019introduction}. When the observed test statistic falls within the critical region, $p$-value of the test statistic becomes less than $\alpha$. In this case, $H_0$ is rejected. If the $p$-value of the test statistic is greater than $\alpha$, the observed test statistic falls outside the critical region. In that case, $H_0$ will be accepted.

\subsection{\label{sec:GoF_test}Goodness-of-fit Test}

\emph{Goodness-of-fit test} is a statistical procedure to decide whether sample data comes from a particular distribution or not. This is a specific hypothesis testing where\\
$H_0$: Sample data comes from a distribution to be tested.\\
$H_a$: Sample data not from that distribution.

Chi-square ($\chi^2$) test\cite{10.1214/aoms/1177729380}, Kolmogorov–Smirnov test (KS test)~\cite{16e7f618-c06b-3d10-8705-1086b218d827}, and Anderson–Darling test~\cite{10.1214/aoms/1177729437} are some examples of the goodness-of-fit test~\cite{good}. Out of these tests, the last two tests are designed for continuous distributions only. However, the KS test is modified later to consider discrete distributions as well~\cite{45e251e1-e12b-3bce-b1a2-da8f4c1c2382}. The $\chi^2$ test has been found more robust than the KS test for the continuous distributions~\cite{32ba378b-c6d3-36dc-9d3e-0b9814b089a4}.

\section{\label{sec:entropy}Quantum Entropy for QRNG}

The entropy generated by an RNG is the source of randomness for its outcome. The outcomes of an RNG follow a probability distribution, say $P(X)$, where $X$ is the underlying random variable.

\begin{definition}
The \emph{Shannon entropy}~\cite{6773024,6773067} of a discrete random variable $X$ is defined as
\begin{equation}
H(X)=-\sum_xP(X=x)\log P(X=x).
\end{equation}
\end{definition}

The minimum entropy of an RNG is the smallest amount of the entropy that the RNG can produce.

\begin{definition}
\emph{Minimum entropy} $H_{min}$ of raw data generated by a RNG is defined as
\begin{equation}
H_{min}:=-\log \left(\max_xP(X=x)\right),
\end{equation}
where $X$ is the underlying discrete random variable.
\end{definition}
This minimum entropy is used to decide the amount of post-processing required for transforming the raw data generated by an RNG to uniform random numbers.

The Shannon entropy of the probability distribution in Eq.~\eqref{E:distIII} is given by
\begin{widetext}
\allowdisplaybreaks
\begin{align}
H(Q)=&-\sum_{i=1}^Nf(i)\log f(i)\notag\\
=&-\sum_{i=1}^N\frac{e^\frac{\mu Td}{N}-1}{1-e^{-\mu Td}}e^{-i\frac{\mu Td}{N}}\Log{\frac{e^\frac{\mu Td}{N}-1}{1-e^{-\mu Td}}e^{-i\frac{\mu Td}{N}}}\notag\\
=&-\frac{e^\frac{\mu Td}{N}-1}{1-e^{-\mu Td}}\sum_{i=1}^Ne^{-i\frac{\mu Td}{N}}\left[\Log{e^{-i\frac{\mu Td}{N}}}+\Log{\frac{e^\frac{\mu Td}{N}-1}{1-e^{-\mu Td}}}\right]\notag\\
=&\frac{e^\frac{\mu Td}{N}-1}{1-e^{-\mu Td}}\left[\Log{e^\frac{\mu Td}{N}}\sum_{i=1}^Nie^{-i\frac{\mu Td}{N}}-\Log{\frac{e^\frac{\mu Td}{N}-1}{1-e^{-\mu Td}}}\sum_{i=1}^Ne^{-i\frac{\mu Td}{N}}\right]\notag\\
=&\frac{e^\frac{\mu Td}{N}-1}{1-e^{-\mu Td}}\Bigg[\Log{e^\frac{\mu Td}{N}}\frac{e^{-\frac{\mu Td}{N}}\left[1-(N+1)e^{-\mu Td}+Ne^{-(N+1)\frac{\mu Td}{N}}\right]}{(1-e^{-\frac{\mu Td}{N}})^2}\notag\\
&\;-\Log{\frac{e^\frac{\mu Td}{N}-1}{1-e^{-\mu Td}}}\frac{e^{-\frac{\mu Td}{N}}(1-e^{-\mu Td})}{1-e^{-\frac{\mu Td}{N}}}\Bigg]\notag\\
=&\frac{1-(N+1)e^{-\mu Td}+Ne^{-(N+1)\frac{\mu Td}{N}}}{(1-e^{-\mu Td})(1-e^{-\frac{\mu Td}{N}})}\Log{e^\frac{\mu Td}{N}}-\Log{\frac{e^\frac{\mu Td}{N}-1}{1-e^{-\mu Td}}}.\label{E:ent_act}
\end{align}
\end{widetext}
This is the entropy generated by the photon arrival-time-based QRNGs we have discussed in this paper.

\section{\label{sec:SPQRNG}Randomness Testing Suites}

Randomness testing is very important to ensure whether an RNG is producing perfect random numbers or not. It tells us about the uniformity of the underlying distribution that the outputs of an RNG follow.
\begin{definition}
Let $S_k=\{0, 1\}^k$ be the set of all $k$-bit binary strings containing $2^k$ elements. A $n$-bit string, $X=x_1x_2\dots x_n,\ x_i\in\{0,1\}$ for all $i$, is called \emph{perfect (or uniformly) random} if and only if for any $k$-bit ($k\leq n$) sub-string $x=x_jx_{j+1}\dots x_{j+k}$, for some $j\leq n-k$, of this string, the probability that $x=y$ for any $y\in S_k$ is $\frac{1}{2^k}$. That is,
\begin{equation}
\begin{aligned}
&X=x_1x_2\dots x_n,x_i\in\{0,1\}\ \forall i,\text{ is true random}\\
\iff&\forall k<n,\forall j\leq n-k\text{ and }\forall y\in S_k,\;P(x=y)=\frac{1}{2^k},
\end{aligned}\notag
\end{equation}
where $x=x_jx_{j+1}\dots x_{j+k}$.
\end{definition}

This definition of a perfect random number uses probability over all possible combinations of the bit string to ensure randomness. However, in a practical case, when we have only a finite sample, this probability cannot be perfectly calculated. Hence this definition cannot be directly used to test the randomness of random numbers.

To overcome this limitation, several randomness testing suites like NIST~\cite{10.5555/2206233}, Dieharder~\cite{Die}, AIS-$31$~\cite{killmann2011proposal}, ENT~\cite{ENT} etc. have been proposed. Recently, Foreman \emph{et al.} also suggested some test suite~\cite{foreman2024} for testing randomness. Each of these suits consists of multiple tests to decide the randomness of the input data. For each test, they calculate the value of the test statistic, and depending on these values the randomness is decided. NIST and Dieharder perform hypothesis testing with the null hypothesis that \emph{the number is random}, AIS-31 performs interval estimation, and ENT directly returns the test statistic values. These suites consider different aspects, such as entropy, serial correlation, frequency of values, repetition of similar patterns etc., to perform the test. Some tests also perform Monte Carlo simulation~\cite{MonteCarlo} to validate the randomness.

\end{document}